%% file: main.tex
\documentclass[journal]{IEEEtran}
\usepackage[hidelinks]{hyperref}
\usepackage{cite}

\usepackage{tikz}
\usetikzlibrary{calc}
\usetikzlibrary{decorations.pathreplacing}
\usetikzlibrary{positioning}
\usepackage{graphicx}
\usepackage{booktabs}
\usepackage{multirow}
\usepackage{xcolor}
\ifCLASSINFOpdf
\else
\fi
\usepackage{amsmath}
\usepackage{amssymb}
\usepackage[caption=false,font=footnotesize]{subfig}
\DeclareMathOperator*{\argmin}{argmin}
\newtheorem{problem}{Problem}

\begin{document}
%
\title{Road User Position Prediction\\ in Urban Environments \\via Locally Weighted Learning$^*$}
%
%
%

\author{Angelos Toytziaridis$^{1}$, Paolo Falcone$^{2}$, Jonas Sj\"oberg$^{3}$
\thanks{*This work was supported by Vinnova FFI project "5G for Connected Autonomous Vehicles in Complex Urban Environments", reference number 2018-05005.}
\thanks{$^{1,2,3}$ All authors are with the Mechatronics group at the Department of Electrical Engineering, Chalmers University of Technology, Gothenburg, Sweden. {\tt\small \{angelos.toytziaridis, paolo.falcone, jonas.sjoberg\}@ chalmers.se}}%
\thanks{$^{2}$ Paolo Falcone is with Engineering Department "Enzo Ferrari", University of Modena and Reggio Emilia, Italy. {\tt\small falcone@unimore.it}}%
}

%
%

\markboth{Journal of \LaTeX\ Class Files,~Vol.~14, No.~8, August~2015}%
{Shell \MakeLowercase{\textit{et al.}}: Bare Demo of IEEEtran.cls for IEEE Journals}
%



\maketitle

\input{abstract}


%
\IEEEpeerreviewmaketitle

\input{introduction}
\input{methodology}
\input{experiments}
\input{discussion}
\input{conclusion}
\input{appendix}



\section*{Acknowledgment}
The authors would like to thank the creators of the freely available python libraries Numpy, Pandas, Matplotlib and Seaborn.

\ifCLASSOPTIONcaptionsoff
  \newpage
\fi



%
\bibliographystyle{IEEEtran}
\bibliography{IEEEabrv,references}

%

\begin{IEEEbiography}{Michael Shell}
Biography text here.
\end{IEEEbiography}

\begin{IEEEbiographynophoto}{John Doe}
Biography text here.
\end{IEEEbiographynophoto}


\begin{IEEEbiographynophoto}{Jane Doe}
Biography text here.
\end{IEEEbiographynophoto}




\end{document}

%% file: abstract.tex
\begin{abstract}
This paper focuses on the problem of predicting the future position of a target road user given its current state, consisting of position and velocity. A weighted average approach is adopted, where the weights are determined from data containing the state trajectories of previously observed road users. 
In particular, a \emph{similarity function} is introduced to extract from data those previously observed road users' states that are most similar to the target's one.
This formulation results in an easily interpretable model with few parameters to calibrate.  The performance of this weighted average model(WAM) is evaluated on the same real-world data as state-of-the-art methods, showing promising results. WAM outperforms the baseline constant velocity model at longer prediction horizons, making WAM suitable for motion planning applications. WAM and a baseline neural network model performs comparably. Still, WAM has only three parameters which are easily interpretable, while the complex neural network model has thousands of parameters which are difficult to analyze.

\end{abstract}

%% file: introduction.tex
\section{INTRODUCTION}
\begin{figure}
	\centering
	\begin{tikzpicture}[scale = 0.4]
		\draw[fill] (0,0)  coordinate (A)
				++(2.5,0) coordinate (B)
				++(2.5,0) coordinate (C)
				++(2.5,0) coordinate (D)
				++(2.5,0) coordinate(E)
				++(2.5,0)  coordinate(F)
				++(2.5,0) coordinate(G);
		\foreach \x in {A, B, C,F}
		{	
			\node[xscale = -1, scale = 0.06,opacity=0.4] at(\x)  {\includegraphics{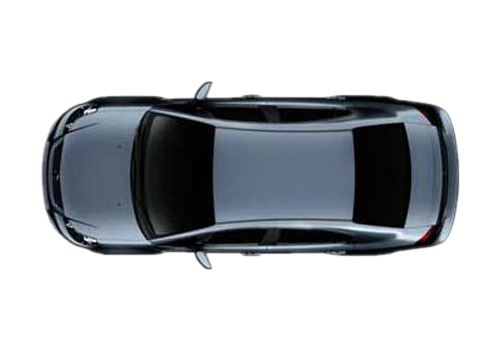}};
			\node[xscale = -1, scale = 0.035,opacity=0.4,rotate=90] at ($(\x)+(0,-1)$) {\includegraphics{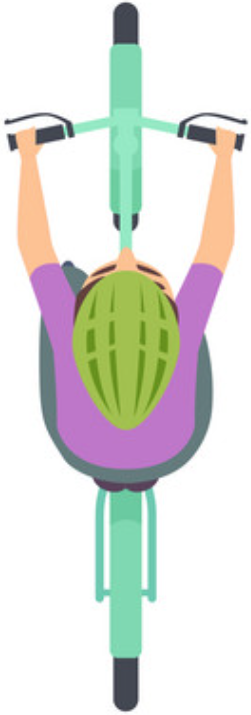}};	
			\node[xscale = -1, scale = 0.2,opacity=0.4] at ($(\x)+(0,-2)$) {\includegraphics{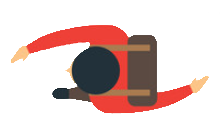}};
			
		}
		\node[xscale = -1, scale = 0.06,opacity=1] at (D) {\includegraphics{figures/car.png}};	
		\node[xscale = -1, scale = 0.035,opacity=1,rotate=90] at ($(D)+(0,-1)$) {\includegraphics{figures/bicycle.png}};
		\node[xscale = -1, scale = 0.2,opacity=1] at ($(D)+(0,-2)$) {\includegraphics{figures/pedestrian.png}};

		\draw (D)++(5,6) node[rotate = 90] (car){
		\includegraphics[scale = 0.1]{figures/car.png}};
		\draw[-latex,thick] (car.center)++(-6,0.5)node[above]{Intelligent vehicle}-- ($(car.center)+(-1.25,0)$) ;
		
		\draw[decorate,decoration={brace,amplitude=5pt}] ($(A)+(-0.5,+1)$)--($(D)+(0.5,1)$)coordinate[midway] (observe);
		\draw ($(car) + (0,-2)$) --  ($(observe)+(0,0.45)$)  node[midway, left= 0.5cm] {past observations};
		\coordinate (predict) at ($(F)+(0,1)$);
		\draw ($(car) + (0,-2)$) --  ($(predict)+(0,0.45)$)  node[midway, right] {predict future};
	\end{tikzpicture}
	\caption{Prediction is required for road safety.}
	\label{fig:prediction_problem}
\end{figure}

\IEEEPARstart{F}{atal} accidents in traffic is the leading cause of death among young children and adults in the world according to World Health Organization\cite{who2018}. A strategy to reduce the number of deaths is to make vehicles intelligent by providing them with computers, sensors, and algorithms for safer navigation; either by helping a human driver or by autonomously driving the vehicle.  
For an intelligent vehicle to safely navigate, or help navigate, without collision it must predict the future state of the environment based on measurements from sensors, as shown in \autoref{fig:prediction_problem}.  

\begin{figure}
	\includegraphics[width=\columnwidth]{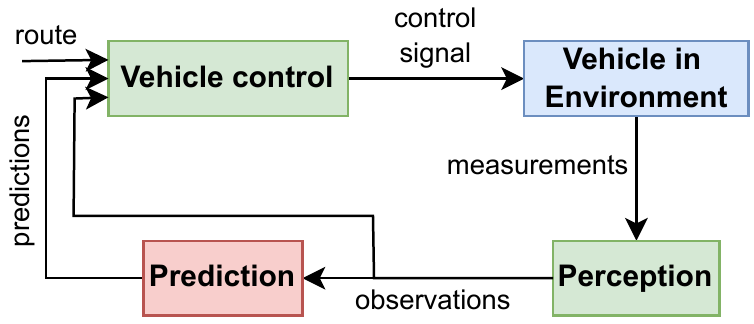}
	\caption{A model of an intelligent vehicle. }
	\label{fig:block_diagram_over_intelligent_vehicle}
\end{figure}

In this work is formulated a method for predicting the future position of a road user given observations. A prediction method of an intelligent vehicle corresponds to the red-colored block in \autoref{fig:block_diagram_over_intelligent_vehicle}. The block diagram illustrates our modular perspective of the problem of an intelligent vehicle navigating in traffic, contrasting the end-to-end approach taken in \cite{bojarski2016}. Red and green blocks can be engineered, while the blue block is phyiscal reality reacting to the control signal. A desired route is recieved by the vehicle control unit, e.g. \cite{batkovic2022enabling}, along with observations of the state of the intelligent vehicle and of surrounding road users, and with predictions of future states of the surrounding road users. A control signal, e.g. acceleration of the vehicle, is determined. The control signal affects the physical world consisting of the vehicle itself and its environment, resulting in the vehicle and environment assuming a state that is percieved by the sensors of the perception unit. Sensor measurements are filtered, fused, and processed, resulting in observations, i.e. collections of higher level information such as position and velocity of a specific road user, \cite{van2018autonomous}. 

An idea to prediction is to assume that road users(RUs), tend to track certain paths, like pedestrians tracking the center of a cross walk, \cite{batkovic2018}. More complex paths can be difficult to determine, hence the idea of formulating prediction as a path planning problem, \cite{karasev2016,rehder2018}, where reference paths are implicitly expressed as a cost function encoding where a RU tends to walk and a destination. An alternative to these reference tracking ideas is to assume that RUs move by switching between operational modes, for exampel move with piecewise constant velocity, and design a method for switching between modes based on current observations. These modes can either be given apriori~{\cite{kooij2019}--\hspace{0.5pt}\cite{minguez2018}} or learned from data \cite{chen2016}. A related idea is to split the prediction problem into  classifying which RUs have critical intentions, such as a pedestrian intending to cross the road, and then predicting critical RUs' future positions~\cite{volz2018}. Other researchers~\cite{koschi2020} propose to model RUs with simpler dynamical models such that there exists inputs that induce the same motion as more complex models. They generate from a simple RU model its set of reachable future positions, from which can be distinguished any critical position.  An approach for constructing complex models of RUs is to start with a flexible black box model and impose structure to force the model to express particular properties, such as interactions between RUs \cite{carrasco2021,alahi2016,gupta2018,li2020,cheng2021amnet,cheng2021exploring}.   Alternatively, a \mbox{model-free} method,~\cite{toytziaridis2019}, is where predictions of the motion of a RU is given by counting in historical observations how many times different velocities occur and then constructing a predicted motion from the most occuring velocities. Another model-free method is $K$-nearest neighbor regression, which is an instance of the Nadaraya-Watson estimator \cite{nadaraya1964} from a class of methods called Locally Weighted Learning \cite{atkeson1997}. A Nadaraya-Watson estimator is a weighted average with weights having a particular structure. In \cite{martin2020}  this structure is parametrized to describe how a target pedestrian's current motion relative to another pedestrian and a destination, relate to the target pedestrian's position at the next sampling instance. By repeatedly evaluating this relation predictions for longer prediction horizons are produced. Their model is evaluated on data of a single pedestrian interacting with a stationary obstacle in a controlled environment. 

In this work prediction of a target road user's future position given its current position and velocity is formulated as a Nadaraya-Watson estimator, i.e. a weighted average, which results in an interpretable model with few parameters. The main contributions are:
\begin{itemize}
	\item This is the first work on car, pedestrian, and bicycle position prediction to apply a Nadaraya-Watson estimator on real traffic data, the Intersection Drone(InD) dataset \cite{inD}.
	\item A formulation of the road user position prediction problem where a Nadaraya-Watson estimator emerges naturally from simple assumptions.
	\item A quantative comparison between our formulation and state-of-the-arts.
	\item An example of how our formulation can leverage interaction between the target road user and the ego vehicle for performing predictions.
	\item An example explaining our method's behavior when deterministic prediction is ill-posed, i.e. when the current state of the target road user relates to two significantly different future states.
\end{itemize}

%% file: methodology.tex
\section{Methodology}\label{sec:method}
\subsection{Position prediction as a weighted average} \label{sec:method:model}
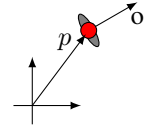
\begin{figure}
	\centering
	\begin{tikzpicture}
		\draw[-{latex}] (-0.25,0) -- (0.65,0);
		\draw[-{latex}] (0,-0.25) -- (0,0.65);
		\coordinate (ped) at (0.75,1);
		\begin{scope}[rotate= 30,shift= {(ped)}]
			\draw[fill=gray] (0,0) ellipse (2pt and 7pt);
			\draw[-{latex}] (0.1,0) --(0.75,0) node[below]{o};
		\end{scope}
		
		\draw[fill=red] (ped) circle (3pt);
		\draw[-{latex}] (0,0) -- ($(0,0) !0.95! (ped)$) node[xshift = -8,yshift=-3]{$p$};
		
 	\end{tikzpicture}
 	\caption{The position $p$ of a road user, e.g. a pedestrian, is expressed w.r.t. an intertial frame. The orientation $o$ of a road user is the unit vector defining the pedestrians forward facing direction.}
 	\label{fig:methodology:orientation}
\end{figure}

\begin{figure}
    \centering
    \includegraphics{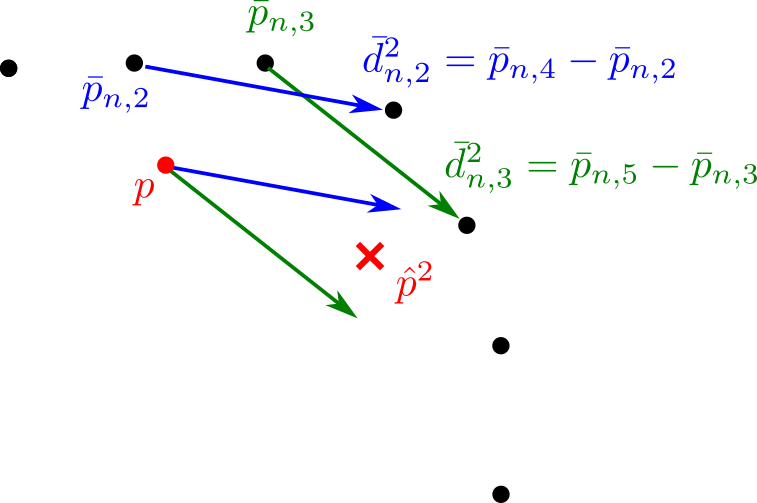}
    \caption{Previously observed road user's positions are shown with black dots. The current position $p$ and the $2$-step predicted position $\hat{p}^2$ of a target road user is shown with a red dot and cross, respectively. Positions $\bar{p}_{n,2}$ and $\bar{p}_{n,3}$ are the positions most similar to position $p$, since they are closest to $p$. Therefore, in this example, the prediction $\hat{p}^2$ is the position closest to position $p+\bar{d}^2_{n,2}$ and position $p+\bar{d}^2_{n,3}$.}
    \label{fig:methodology:explain_model}
\end{figure}

Let the current state of a \emph{target road user} be~${R=(p,s,o)}$, with $p\in\mathbb{R}^{2}$ the position in an inertial reference frame, ${s\in\mathbb{R}}$ the speed and $o\in\mathbb{R}^2$ the orientation. 
The orientation is the unit vector whose direction defines the forward direction of the road user, as shown in \autoref{fig:methodology:orientation}.
Assume that a collection~$\mathcal{C}=\{\mathcal{C}_n\}_{n=1}^N$ of 
previously observed trajectories $\mathcal{C}_n = \{\bar{R}_{n,t}\}_{t=1}^{T_n}$ of road users' states are available, where~$\bar{R}_{n,t}=(\bar{p}_{n,t},\bar{s}_{n,t},\bar{o}_{n,t})$ denotes the $t$-th sample of the $n$-th trajectory. 
The road user's position prediction problem considered in this paper is stated as:
\begin{problem}\label{prob:prediction}
Given a target road user's current state~$R$ and a collection~$\mathcal{C}$ of previously observed road users' state trajectories, determine a prediction $\hat{p}^H$ of the target road user's position at $H$ sampling instants later, where $H$ is called the \emph{prediction horizon}.
\end{problem}
%
The approach adopted in this paper to solve Problem~\ref{prob:prediction} relies on the observation that if~$R$ is \emph{similar}, in some sense, to the sample $\bar{R}_{n,t}$, then a prediction $\hat{p}^H$ is likely to be \emph{similar} to the position $p+(\bar{p}_{n,t+H}-\bar{p}_{n,t})$. \autoref{fig:methodology:explain_model} illustrates the construction of a prediction based on this observation when similarity between $R$ and $\bar{R}_{n,t}$ is measured by the distance between positions $p$ and $\bar{p}_{n,t}$.
Hence, a \emph{similarity function}, taking the current state $R$ of a target road user and a sample $\bar{R}_{n,t}$ and relating them to a real number representing similarity between $R$ and $\bar{R}_{n,t}$ is to be introduced. In this paper a similarity function is built upon the Euclidean distance between $p$ and $\bar{p}_{n,t}$, the difference between speeds $s$ and~$\bar{s}_{n,t}$, and between the orientations $o$ and $\bar{o}_{n,t}$, and is defined in \autoref{sec:method:defining_similarity}.

%
Call the difference $\bar{p}_{n,t+H}-\bar{p}_{n,t}$ an \emph{$H$-step displacement} and denote it by $\bar{d}^H_{n,t}$ and introduce the $H$-step predicted displacement $\hat{d}^H$ such that the $H$-step ahead predicted position~$\hat{p}^H$ is calculated as 
\begin{equation}
	\hat{p}^H = p + \hat{d}^H,
\end{equation}
where $\hat{d}^H$ is closer to $\bar{d}_{n,t}^H$ the more similar $R$ is to $\bar{R}_{n,t}$. 
Once the similarity between $R$ and $\bar{R}_{n,t}$ is expressed by a function~$\sigma(R,\bar{R}_{n,t})$, that is non-negative and not zero everywhere, the displacement $\hat{d}^H$ is defined as the member of the singleton
\begin{equation}\label{eq:methodology:optimality_formulation}
		\argmin_{\hat{d}^H} \quad \sum_n\sum_t \sigma(R,\bar{R}_{n,t})\cdot\|\hat{d}^H-\bar{d}_{n,t}^H\|^2.
\end{equation}
It is shown in Appendix~\ref{sec:appendix:proof} that the $H$-step predicted displacement~$\hat{d}^H$ is a weighted average of every sampled $H$-step displacement~$\bar{d}^H_{n,t}$:
\begin{equation}\label{eq:methodology:nadaraya-watson}
	\hat{d}^H = \sum_n \sum_t \frac{\sigma(R,\bar{R}_{n,t})}{\sum_m \sum_\tau \sigma(R,\bar{R}_{m,\tau})}\bar{d}^H_{n,t}.
\end{equation}
\autoref{eq:methodology:nadaraya-watson} is recognized as a case of the Nadaraya-Watson estimator \cite[Sec. 2.8.2]{hastie2009elements}.  

In this section a prediction model was formulated that takes as input the current position, speed and orientation of a target road user and returns as output the target road user's future position. One may use the same methodology to define more sophisticated models, for example that includes as input the positions of road users surrounding the target road user, see \autoref{sec:discussion:interaction}.

\subsection{Defining similarity between road users}\label{sec:method:defining_similarity}

	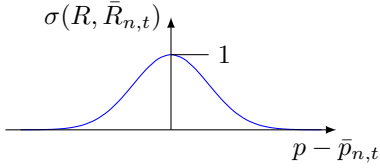
\begin{figure}
	    \centering
	    \begin{tikzpicture}
	        \draw[scale=1, domain=-2:2, smooth, variable=\x, blue] plot ({\x}, {exp(-2*\x*\x)});
	        \draw[-{latex}] (-2.2,0)--(2.2,0) node[below]{$p-\bar{p}_{n,t}$};
	        \draw[-{latex}] (0,0)--(0,1.5) node[left] {$\sigma(R,\bar{R}_{n,t})$};
	        \draw[shift ={(0,1)}] (0,0)--(0.5,0) node[right] {$1$};
	    \end{tikzpicture}
	    \caption{The similarity $\sigma$ as a function of the difference between the current position $p$ of a target road user and the position $\bar{p}_{n,t}$ of a previously observed road user from the dataset.}
	    \label{fig:methodology:explain_similarity}
	\end{figure}

	The example in \autoref{fig:methodology:explain_model} suggests that the smaller the distance between the positions $p$ and $\bar{p}_{n,t}$ is, the higher the similarity $\sigma(R,\bar{R}_{n,t})$ should be. For example, for a similarity function only depending on the distance between positions one could define the function~$\sigma$ as illustrated in \autoref{fig:methodology:explain_similarity}. For a general state~$R$, the similarity is assumed to monotonically decrease with the difference in position~${\|p-\bar{p}_{n,t}\|}$, in speed~${\|s-\bar{s}_{n,t}\|}$, and in orientation, defined as the angle~$\bar{\theta}_{n,t}$ between the orientations $o$ and $\bar{o}_{n,t}$. In this paper the similarity function is formally defined using an instance of the Gaussian kernel~\cite{atkeson1997}: 
	\begin{equation}\label{eq:similarity_metric}
			\sigma(R,\bar{R}_{n,t}) = \exp(-(
			a\|p-\bar{p}_{n,t}\|^2 
			+ b|s-\bar{s}_{n,t}|^2 
			+ c|\bar{\theta}_{n,t}|^2)),
	\end{equation}
	with $a,~b,~c\in\mathbb{R}^{+}$ being parameters that 
	shape the similarity function. The definition~\eqref{eq:similarity_metric} implies that restrictions of the function~$\sigma(R,\bar{R}_{n,t})$ such that any two of~$p-\bar{p}_{n,t}$,~$s-\bar{s}_{n,t}$, and~$\bar{\theta}_{n,t}$ are set to fixed values and the third is varied, look like the function in \autoref{fig:methodology:explain_similarity}.

\subsection{Learning the parameters in the similarity function} \label{sec:method:identification}
	In the similarity function defined in \mbox{\autoref{sec:method:defining_similarity}} the parameters $a,~b,~c$ has to be learnt from available data. 
	To this aim, define the set
	\begin{equation}
		C = \bigcup_n \bigcup_t \{(\bar{R}_{n,t},\bar{d}_{n,t}^H)\}.
	\end{equation}
	of pairs of an observed road user state and a displacement. The set~$C$ is such that if~$(\bar{R},\bar{d}^H)\in C$ then~${\bar{R} = (\bar{p},\bar{s},\bar{o})}$ is a sample of a road user's state and~$\bar{p}+\bar{d}^H$ is the sample of the road user's position at $H$ sampling instants later.
	Using set~$C$, the $H$-step predicted displacement~$\hat{d}^H$ in~\eqref{eq:methodology:nadaraya-watson} can be expressed as
	\begin{equation} \label{eq:methodology:nadaraya-watson_alternative}
		\hat{d}^H(R, C) = 	
		\sum_{(\bar{R}_1,\bar{d}^H_1)\in C} \frac{\sigma(R,\bar{R}_1)}
		{\displaystyle\sum_{(\bar{R}_2,\bar{d}^H_2)\in C} \sigma(R,\bar{R}_2)}\bar{d}_1^H.
	\end{equation}
	A possible parameter learning formulation is
	\begin{equation} \label{eq:methodology:classic_identification}
		\argmin_{a,b,c} \frac{1}{|C|}\sum_{(\bar{R},\bar{d}^H)\in C} \|\bar{d}^H - \hat{d}^H(\bar{R}, C)\|^2
	\end{equation}
	whose minimum occurs when
	\begin{equation}
		\sigma(R,\bar{R}) = 
		\begin{cases}
			1, \text{if } R = \bar{R} \\
			0, \text{otherwise}
		\end{cases}
	\end{equation}
	which is attained as $a~,b~,c$ tend to infinity, since then the Gaussian kernel \eqref{eq:similarity_metric} collapses. To avoid this a learning  formulation must be such that for some $\bar{R}_1$, 
	if $R=\bar{R}_1$ in \eqref{eq:methodology:nadaraya-watson_alternative}, then $\bar{d}^H_1$ in \eqref{eq:methodology:nadaraya-watson_alternative} is different from $\bar{d}^H$ in \eqref{eq:methodology:classic_identification}.
	This is avoided by using $K$-fold cross-validation \cite[Sec. 7.10]{hastie2009elements}, where the set $C$ is partitioned in subsets~${C_1, C_2,\dots, C_K}$ such that:
	\begin{enumerate}
	    \item $C_i\cap C_j=\emptyset$ when $i\neq j$,
	    \item $\bigcup_{k=1}^K C_k = C$,
	    \item $\forall i,j, |C_i|=|C_j|$,
	    \item and that elements in $C$ stemming from the same trajectory belong to the same subset.
	\end{enumerate}
	The last constraint is formally expressed as: for every $k$ if~${(\bar{R}_1,\bar{d}^H_1),(\bar{R}_2,\bar{d}^H_2)\in C_k}$, then there exists a trajectory in the collection $\mathcal{C}$ with index $n$ and two time indices $t_1$ and $t_2$ such that $(\mathcal{C}_n)_{t_1} =\bar{R}_1$ and $(\mathcal{C}_n)_{t_2} =\bar{R}_2$.  This constraint is imposed to avoid that data on which the prediction model is evaluated depends causally on the data used by the prediction model to calculate a predicted displacement.
    In~\mbox{$K$-fold cross-validation} the fitting score of a combination of parameters on subset $C_k$ is defined as
	\begin{equation}
		L_k =  \frac{1}{|C_k|}\sum_{(\bar{R},\bar{d}^H)\in C_k} \|\bar{d}^H - \hat{d}^H(\bar{R}, \bigcup_{\substack{j=1\\j\neq k}}^K C_j)\|^2
	\end{equation}
	and  optimal parameters are determined by 
	\begin{equation} \label{eq:method:learning_problem}
		\begin{aligned}
			&\argmin_{a,b,c} & & \frac{1}{K}\sum_{k=1}^K L_k.
		\end{aligned}
	\end{equation}

%% file: experiments.tex
\section{Experiments}

A dataset with real world traffic data, with three types of road users at two different intersections is used for evaluating the model formulation in \autoref{sec:method}. For every combination of type and intersection a model is implemented, hence, in total six models are evaluated.  A portion of the dataset is used for fitting parameters of a model and the remaining data is used for assessing the performance of the fitted model. 

	\subsection{Implementation details} \label{sec:experiment:implementation}
	    The larger collection~$\mathcal{C}$ in \autoref{sec:method} is, the longer is the time required to compute the $H$-step predicted displacement~$\hat{d}^H$. To alleviate the computational burden, the definition of similarity in \autoref{sec:method:defining_similarity} is modified such that if the distance~${\|p-\bar{p}_{n,t}\|}$ is larger than $r=15$ meters, then the similarity~$\sigma(R,\bar{R}_{n,t})$ is zero. This has negligible impact on the results since the function~$\exp(-x)$ decays quickly anyway. Furthermore, to efficiently find $\bar{R}_{n,t}$ in the data corresponding to non-zero similarity it is recommended to use a balltree data structure.
	    
	     It  customary to let algorithms depend on samples that are at most $2.8$ seconds old, \cite{carrasco2021,alahi2016,gupta2018,cheng2021amnet,cheng2021exploring}. The formulation in \autoref{sec:method} can be re-formulated such that $R$ and $\bar{R}_{n,t}$ includes such states, but this is not explored in this paper. Still, to enable algorithms depending on such samples to be evaluated on the same data as the algorithm in this paper is evaluated on, a constant $c$ is introduced such that $c$ times the sampling time is $2.8$ seconds, and the sample index $t$ in section \autoref{sec:method} is restricted to $t\geq c$ rather than $t\geq 1$.
		
	\subsection{Choice of dataset and data pre-processing}\label{sec:experiments:preprocessing}
		The weighted average model in \autoref{sec:method} is tested on  the Intersection Drone(InD) dataset\cite{inD}, which consists of trajectories describing the motion of four types of road users: \emph{pedestrians}, \emph{bicyclists}, including motorcyclists, \emph{cars} and \emph{trucks/buses}. The trajectories have been extracted from recordings of four urban intersections. Although the dataset is constructed without human annotation, the InD dataset is chosen because of its large size and its convenient location dependent reference frame.

		To reduce computational burden every trajectory is downsampled from $25$ Hz to $2.5$ Hz. The longest prediction horizon adopted in this paper is $4.8$ seconds, as is customary in the literature \cite{carrasco2021,alahi2016,gupta2018,cheng2021amnet,cheng2021exploring}. Combining this with that algorithms typically depend on at most $2.8$ seconds old samples, see \autoref{sec:experiment:implementation}, means that trajectories shorter than $2.8+4.8$ seconds are too short for applying prediction. Hence, too short trajectories are discarded. 
		It is assumed that a pedestrian or a bicycle trajectory with a speed larger than $15$ km/h and $35$ km/h, respectively, is an outlier and is therefore discarded.
		For every road user, the dataset includes a velocity trajectory $\{v_t\}$, from which is constructed an orientation trajectory~$\{o_t\}$ such that if the velocity~$v_t$ is non-zero, then~${o_t=v_t/\|v_t\|}$, otherwise~$o_t$ is equal to the most previous non-zero normalized velocity.
		To further limit the required computations, trajectories corresponding to a road user being mostly stationary are discarded, where a mostly stationary trajectory is defined as the 95 percentile of the speeds along the trajectory being less than 0.36 km/h. Alternatively, one could partly trim stationary parts of a trajectory, hence keeping all of the more interesting parts. The dataset contains few observations of the road user type truck/bus, therefore the types truck/bus and car are merged into a single type called \emph{vehicle}. \autoref{tab:data_inspection:numberRoadUsers} shows the number of road users per type and location before and after processing the data.
		
		It should be pointed out that the model in \autoref{sec:method} performs poorly at regions in the dataset that have few road users traversing it.
		Hence, due to the limited amount of data available at locations 3 and 4 these locations are discarded. 
		Furthermore, note that \autoref{fig:data:locii_of_trajs_location1} and \autoref{fig:data:locii_of_trajs_location2} indicate that in some regions few road users have been observed, consequentially we expect worse results in those regions.
		
\begin{figure}
		\centering
		\foreach \class in {vehicle,bicycle,pedestrian}
		{
			\subfloat[\class]{
				\centering
				\input{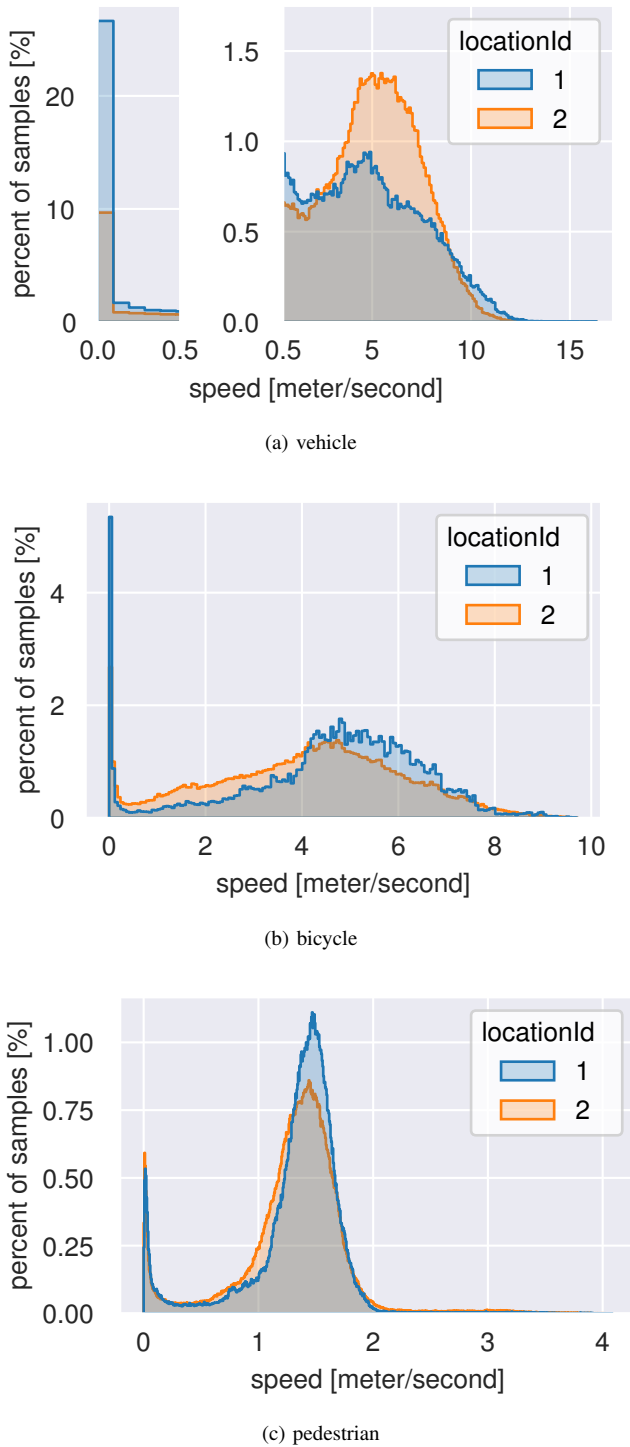}

				\label{fig:data:histograms_of_speed_\class}
			}%
		}%
		\caption{Histograms showing the distributions of speeds in the processed dataset.}
		\label{fig:data:histograms_of_speed}
	\end{figure}

		\autoref{fig:data:histograms_of_speed} shows an over-representation of speeds close to zero. A close inspection of the dataset indicates that this is because of some trajectories being partly
	 	stationary; for example, a pedestrian standing still at a location or a vehicle stopping at the intersection to give way to another vehicle.
		
		For every type of road user, and for every location, the subset of data containing only trajectories from road users of a type at a location is selected, then the subset is split into training and test data. Care must is taken when splitting the data, since some splits can introduce causal relationships between training and test data. For example, consider two friends walking side by side, if one of these pedestrians is assigned as training data and the other as test data, then the future position of the pedestrian in the test data is well-explained by the data from the other pedestrian in the training data. Therefore, since the InD dataset consists of~$33$ drone recordings, the data is split such that some of the recordings are designated as training data. Sticking to the rule-of-thumb that~$70$~\% of data should be training data, the data is split such that for every combination of road user type and location the portion of training data is as close as possible to~$70$~\%. This is ensured using an exhaustive brute force search. The training and test data corresponding to a road user type~$\tau$ and location~$L$ are denoted by the collections~$\mathcal{C}^\text{train}_{\tau,L}$ and~$\mathcal{C}^\text{test}_{\tau,L}$, respectively, defined as in \autoref{sec:method:model}.

	\input{tables/numberOfRoadUsers}

	\foreach \locationId in {1,2}
	{
	\begin{figure*}
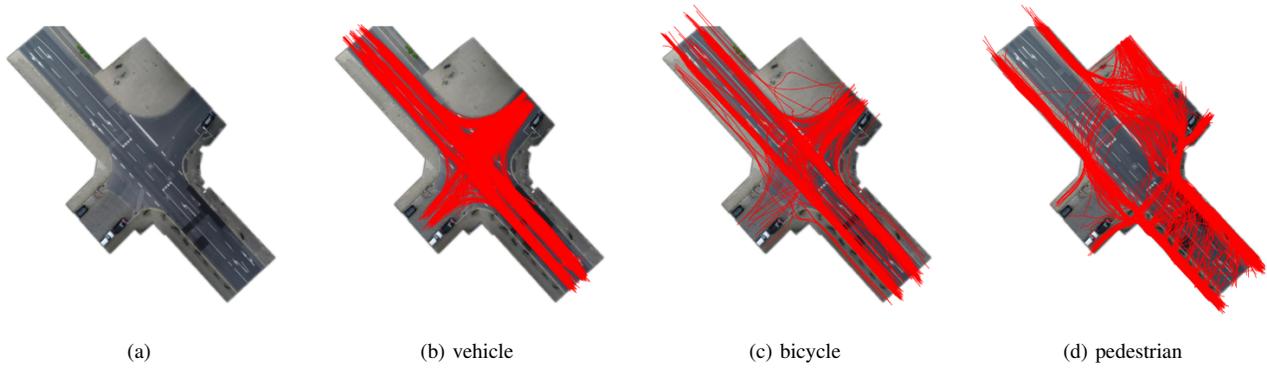

		\centering
		\subfloat[]{
			\includegraphics[angle=0,width=0.22\textwidth]{figures/background_location\locationId.png}
		}%
		\foreach \class in {vehicle,bicycle,pedestrian}
		{
			\subfloat[\class]{
				\includegraphics[angle=0,width=0.22\textwidth]{figures/locii_of_positions_location\locationId_class=\class}
				\label{fig:data:locii_of_trajs_location\locationId_\class}
			}%
		}
		\caption{Every red curve is a trajectory of positions of a road user at location \locationId.}
		\label{fig:data:locii_of_trajs_location\locationId}
	\end{figure*}
	}

		\foreach \class in {vehicle,bicycle,pedestrian}
		{
			\begin{figure*}
				\centering
				\foreach \locationId in {1,2}{
					\subfloat[Location \locationId]{
					\input{figures/boxplot_location=\locationId_class=\class.pgf}
					\label{fig:results:boxplot:location\locationId_\class}
					}
				}
				\caption{Results for \class s. Statistics of $H$-second prediction errors of the weighted average model(WAM) and baseline models. The white lines indicate median values. The lower and upper sides of a rectangle indicate the first and third quartiles, respectively. }
				\label{fig:results:boxplot:\class}
			\end{figure*}
		}	

	\subsection{Learning parameters $a,b,c$ with 5-fold cross-validation}\label{sec:experiment:parameter_identification}
	    Each of three road user types, vehicle, bicycle and pedestrian, in combination with one of the two locations are modeled as in \autoref{sec:method}. For the model of road users of type $\tau$ at location $L$, parameters $a,~b,~c$ in \eqref{eq:similarity_metric} are learned using $5$-fold cross-validation, as explained in \autoref{sec:method:identification}, on the training data~$\mathcal{C}^\text{train}_{\tau,L}$. Ideally the subsets derived from~$\mathcal{C}^\text{train}_{\tau,L}$ should all have equal cardinality, in practice the cardinalities are \emph{very similar}. The learning problem \eqref{eq:method:learning_problem} is approximately solved using grid-search  with manual refinements of the search space.
	    
	    \begin{table}
	    	\centering
	    	\caption{Learned parameters for the weighted average model.}
		    \begin{tabular}{lllll}
		    \toprule
		    & parameter & $a$ & $b$ & $c$\\
		    type & location & & & \\
		    \toprule
		    \multirow[t]{2}{*}{vehicle} & 1 & 0.5 & 1 & 50 \\
		    								  & 2 & 0.5 & 1 & 200 \\
		    \multirow[t]{2}{*}{bicycle} & 1 & 0.5 & 20 & 50 \\
		    								  & 2 & 0.25 & 1 & 100 \\
		    \multirow[t]{2}{*}{pedestrian} & 1 & 0.25 & 20 & 50 \\
		    								  		& 2 & 0.1 & 50 & 50 \\
			\bottomrule
		    \end{tabular}
		    \label{tab:learned_parameters}
	    \end{table}

		Learned parameters are shown in \autoref{tab:learned_parameters}. The large values of~$c$, which relates to the rate of decay if similarity between two orientations of two road users, indicates that a predicted $H$-step displacement is mostly influenced by 
		the previously observed road user states with most similar orientation to the target road user's current state. In~\autoref{sec:discussion:large_parameters} is an explanation of why parameters assume large values.
		
	\subsection{Baseline models: constant velocity and neural network} \label{sec:baseline_models}
		Two baseline models are chosen for comparison, which are detailed next. 
		The baseline models are also trained and evaluated per combination of type of road user and location.
		
	\subsubsection{The constant velocity model}
        For a target road user with position~$p$, speed~$s$, and orientation~$o$, the $H$-step predicted position is for a sampling time~$\Delta$ defined as~$\hat{p}^H=p+soH\Delta$.
	
	\subsubsection{The neural network model}
		For a target road user with current position~$p$, speed~$s$, and orientation~$o$ the $H$-second predicted displacement~$\hat{d}^H$ is the output of a neural network taking as input~$(p,s,o)$, and the predicted position is~${\hat{p}^H = p + \hat{d}^H}$. The neural network has one hidden layer with every neuron having ReLU as activation function. The number of neurons are learned via 5-fold cross-validation as in \autoref{sec:experiment:parameter_identification}, and are shown in \autoref{tab:learned_parameters_NN}. The weights and biases are learned using the default settings in Scikit-Learn, except for the number of maximum iterations being changed to~$400$ from~$200$ to guarantee convergence, and the setting \emph{random\_state} being set to zero to ensure reproducability.
	
	    \begin{table}
	    	\centering
	    	\caption{Learned parameters parameters for the baseline neural network model.}
		    \begin{tabular}{lll}
		    \toprule
		    & parameter & number of neurons \\
		    type & location & \\
		    \toprule
		    \multirow[t]{2}{*}{vehicle} & 1 & 500  \\
		    								  & 2 &  400\\
		    \multirow[t]{2}{*}{bicycle} & 1 & 225 \\
		    								  & 2 & 350\\
		    \multirow[t]{2}{*}{pedestrian} & 1& 500 \\
		    								  		& 2 &  150\\
			\bottomrule
		    \end{tabular}
		    \label{tab:learned_parameters_NN}
	    \end{table}

	\subsection{Validation of the weighted average model}\label{sec:experiments:validation}
		The weighted average model (WAM) from \autoref{sec:method} is validated both with respect to the length of the prediction horizon and to the geometry of the road network. For the former, statistics of distributions of prediction errors at different prediction horizons is studied, while for the latter the prediction horizon is set to 4.8 seconds and prediction errors at different current positions is studied.
		
	    WAM with the parameters in \autoref{tab:learned_parameters} is evaluated on the test datasets $\mathcal{C}^\text{test}_{\tau,L}$ introduced in \autoref{sec:experiments:preprocessing}. 
	    If sample~${\bar{R}_{n,t}^{\text{train}}=(\bar{p}_{n,t},\bar{s}_{n,t},\bar{o}_{n,t})}$ is assumed as the current state of a target road user, then the $H$-step future position of the road user is~$\bar{p}_{n,t+H}$. 
	    Let the $H$-step predicted position be~$\hat{p}^H$, assuming that~$\bar{R}_{n,t}^{\text{train}}$ is the current state of a target road user.
	    Then, the corresponding \emph{$H$-step prediction error} is defined as~$\|\bar{p}_{n,t+H}-\hat{p}^H\|$. In the analysis of test results it is more meaningful to consider the prediction horizon $H$ in seconds. Henceforth, we shall refer to $H$-\emph{second} prediction errors instead.
	    
		Figures~\ref{fig:results:boxplot:vehicle},~\ref{fig:results:boxplot:bicycle},~and~\ref{fig:results:boxplot:pedestrian},  show that at location~1 and location~2, for every road user type, the median and quartile $H$-second prediction error increases as $H$ increases. In general, the all models perform comparably up to about~$1.6$ seconds.  In all cases WAM and neural network model typically outperform the constant velocity model, especially at longer horizons and in reducing large errors.
		\autoref{fig:results:boxplot:vehicle} shows that for vehicles at location~1 WAM performs slightly worse than the neural network model, while at location~2 the performance is comparable.
		\autoref{fig:results:boxplot:bicycle} shows that for bicycles at location~1 WAM and the neural network perform comparably. At location~2 WAM performs better than the neural network at long horizons.
		\autoref{fig:results:boxplot:pedestrian} shows that for pedestrians at location~1  WAM performs slightly worse at short horizons and better at long horizons than the neural nerwork. At location~2 WAM outperforms the neural network model at long horizons.

		At annotation~A in \autoref{fig:experiments:position_vs_4.8seconds_location2_vehicle} the errors are larger than at other locations, although according to \autoref{fig:data:locii_of_trajs_location2_vehicle} there appears to be a lot of data at annotation~A. This indicates that the defined similarity~\eqref{eq:similarity_metric} fails to completely distinguish between the future outcomes \emph{continue}, \emph{turn}, and \emph{stop}; which is reasonable since current position, current speed and current orientation do not uniquely define a future position. Furthermore, the large errors at annotation~A may also be due to the model being unable to predict \emph{when} the target vehicle starts moving after being stationary, as exemplified in \autoref{sec:discussion:interaction}. The same argument can be made for annotation~B in~\autoref{fig:experiments:position_vs_4.8seconds_location2_bicycle}. A trajectory of a pedestrian at annotation~C in~\autoref{fig:experiments:position_vs_4.8seconds_location2_pedestrian} has a large error. If we compare the distance between samples along the trajectory with that of other pedestrian trajectories, then we find that the speed of this pedestrian is about twice that of the other pedestrians. According to \autoref{fig:data:histograms_of_speed_pedestrian} such fast moving pedestrians are few in the data, thus it is expected that WAM performs less well. Similarly, at annotation~D in \autoref{fig:experiments:position_vs_4.8seconds_location2_pedestrian} the large errors of WAM is explained by there being few data in this region, see  \autoref{fig:data:locii_of_trajs_location2_pedestrian}. Not surprisingly, the neural network manages to model outlier trajectories better, thanks to the model having many parameters.

	\begin{figure*}
		\centering
		\foreach \class in {vehicle,bicycle,pedestrian}
		{
			\subfloat[\class]
			{
				\includegraphics[width=\textwidth]{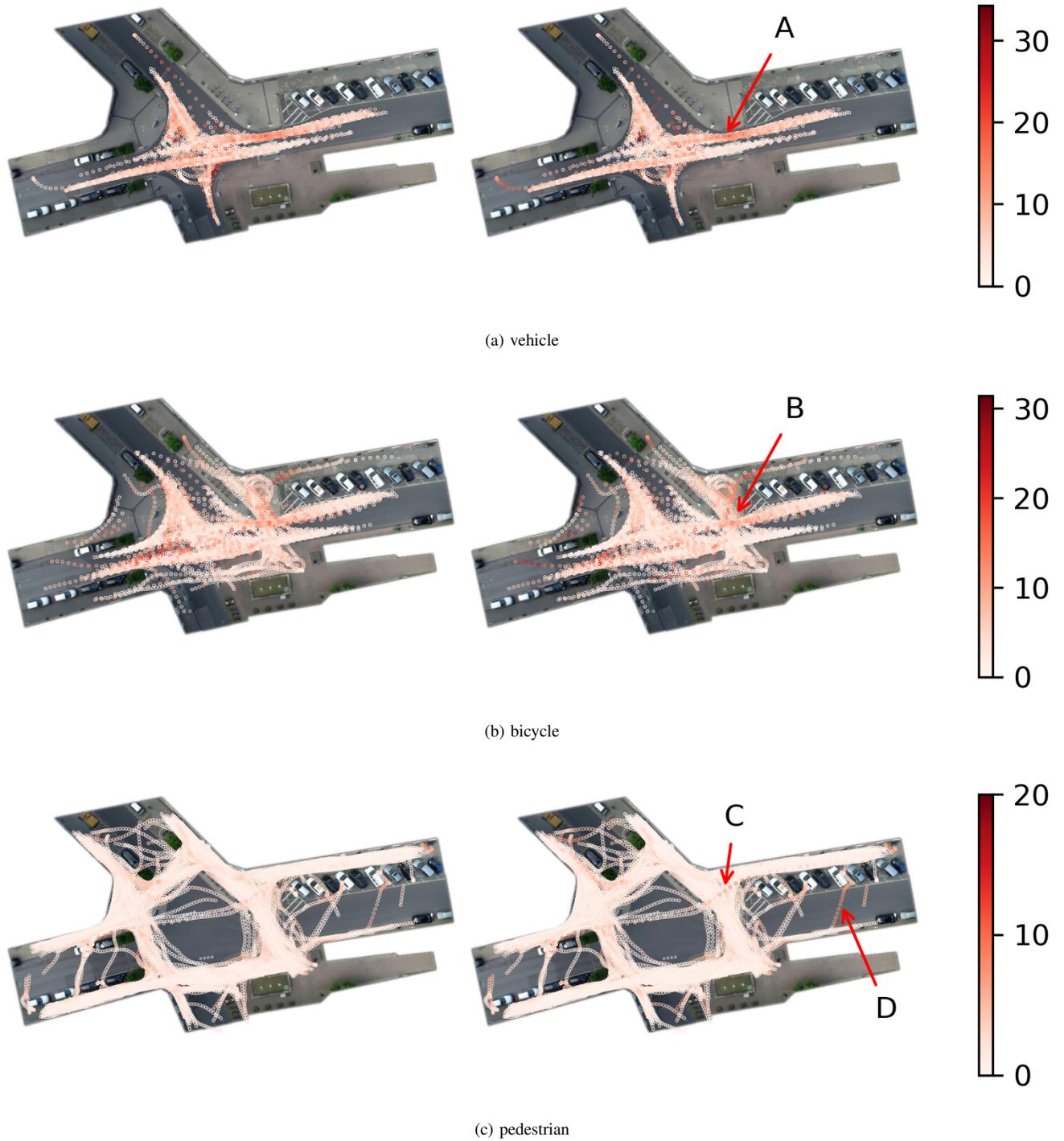}
				\label{fig:experiments:position_vs_4.8seconds_location2_\class}
			}
		}
		\caption{Location 2. Left is the neural network prediction model. Right is the weighted average model. A circle is centered at a sample of the position of a roaduser. The magnitude in meters of a $4.8$-second prediction error of a prediction model is indicated by the color of a circle.}
		\label{fig:experiments:position_vs_4.8seconds_location2}
	\end{figure*}

	\subsection{Comparison with state-of-the-art on another subset of InD}
		In \cite{carrasco2021} is reported state-of-the-arts results on a subset of the inD dataset that is different from the subset defined in \autoref{sec:experiments:preprocessing}. Furthermore, also the training and test split is different. For the given split, a weighted average model from \autoref{sec:method} is defined for every combination of road user type and location. These models are trained like in \autoref{sec:experiment:parameter_identification}, and the corresponding prediction errors on the test set are determined. During these evaluations the constant $r$ in \autoref{sec:experiment:implementation} is set to 100 due to data being sparsely distributed over the intersections at the locations that were discarded in \autoref{sec:experiments:preprocessing}. It is custom in the literature to compare models by their average $4.8$-second prediction error, called the \emph{Final Displacement Error}(FDE), and to compute the average of the $0.4$-second prediction error, $0.8$-second prediction error, $\dots$, $4.8$-second prediction error, and compare the average of these averages, which is called the \emph{Average Displacement Error}(ADE).  \autoref{tab:results:comparison_to_stateoftheart} shows the FDE and ADE calculated per location, i.e. types of road user are not distinguished when computing the averages. In \autoref{tab:results:comparison_to_stateoftheart} the locations are labeled as in \cite{carrasco2021}, where labels B and C correspond to locations 1 and 2, respectively, in this paper. \autoref{tab:results:comparison_to_stateoftheart} shows that the weighted average model of \autoref{sec:method} requires further improvement. A possible improvement is to re-formulate the model such that it includes as input also the position of road users surrounding the target road user, as exemplified in \autoref{sec:discussion:interaction}. In fact, all other models in \autoref{tab:results:comparison_to_stateoftheart}  include such inputs, as well past positions of both the target vehicle and surrounding vehicle, which may explain their superior performance. Notice that out of these more complex algorithms the models \cite{alahi2016} and \cite{gupta2018}, with approximately $8\,500\,000$ and $11\,000$ number of parameters, respectively, actually perform comparable to the weighted average model, which has only $3$ parameters.
		
	\begin{table}
		\centering
		\caption{Comparison with state-of-the-art on ADE /FDE scores. Scores of methods \cite{carrasco2021,alahi2016,gupta2018,li2020,cheng2021amnet,cheng2021exploring} were originally reported in \cite{carrasco2021}.}
		\begin{tabular}{lllll}
			\toprule
			Location &  A & B & C & D\\
			Method  &             &             &             & \\
			\midrule
			S-LSTM \cite{alahi2016}  & 2.29 / 5.33 & 1.28 / 3.19 & 1.78 / 4.24 & 2.17 / 5.11\\
			S-GAN  \cite{gupta2018}  &3.02 / 5.30 & 1.55 / 3.23 & 2.22 / 4.45 & 2.71 / 5.64\\
			GRIP++ \cite{li2020}& 1.65 / 3.65 & 0.94 / 2.06 & 0.59 / 1.41 & 1.94 / 4.46 \\
			AMENet  \cite{cheng2021amnet}& 1.07 / 2.22 & 0.65 / 1.46 & 0.83 / 1.87 &  0.37 / 0.80 \\
			DCENet \cite{cheng2021exploring} & 0.96 / 2.12 & 0.64 / 1.41 &  0.86 / 1.93 &  0.28 / 0.62 \\
			SCOUT  \cite{carrasco2021} & 0.67 / 1.55 & 0.48 / 1.08 & 0.30 / 0.69 & 0.40 / 0.83\\
			\input{tables/rows_for_table_with_state-of-the-art.txt}
			\bottomrule
		\end{tabular}
		\label{tab:results:comparison_to_stateoftheart}
	\end{table}


%% file: tables/numberOfRoadUsers.tex
\begin{table}
\centering
\caption{Number of road users per location and class in original/processed datasets.}
\label{tab:data_inspection:numberRoadUsers}
\begin{tabular}{lllll}
\toprule
locationId &           1 &            2 &           3 &            4 \\
type      &             &              &             &              \\
\midrule
bicycle    &   434 / 360 &  1700 / 1601 &     39 / 20 &      86 / 48 \\
pedestrian &   801 / 755 &  2099 / 2015 &     44 / 42 &    163 / 156 \\
vehicle    &  2503 / 959 &  2436 / 2094 &  1196 / 289 &  2098 / 1442 \\
\bottomrule
\end{tabular}
\end{table}

%% file: tables/rows_for_table_with_state-of-the-art.txt
Const. vel. & 3.51 / 8.92 & 1.22 / 3.11 & 1.86 / 4.56 & 1.55 / 3.96\\
WAM & 2.64 / 5.87 & 1.48 / 3.42 & 1.73 / 3.96 & 2.91 / 6.24\\

%% file: discussion.tex
\section{Discussion} 

	\subsection{Example illustrating why identified parameters are large}\label{sec:discussion:large_parameters}
\begin{figure*}
			\centering
			\subfloat[ ]{
				\includegraphics[width=0.5\textwidth]{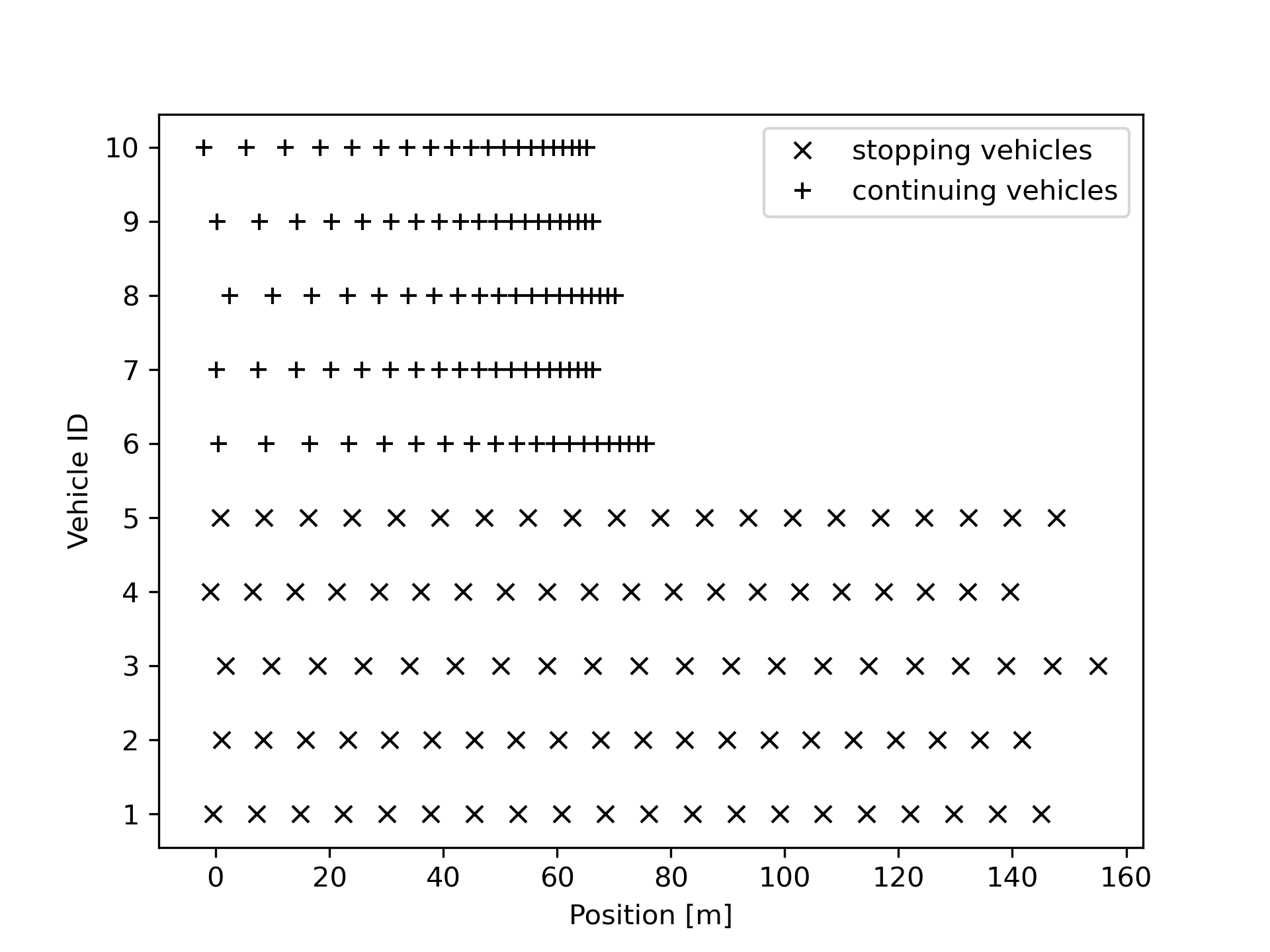}
				\label{fig:discussion:example_explaining_large_parameters_data}
			}
			\subfloat[]{
				\includegraphics[width=0.5\textwidth]{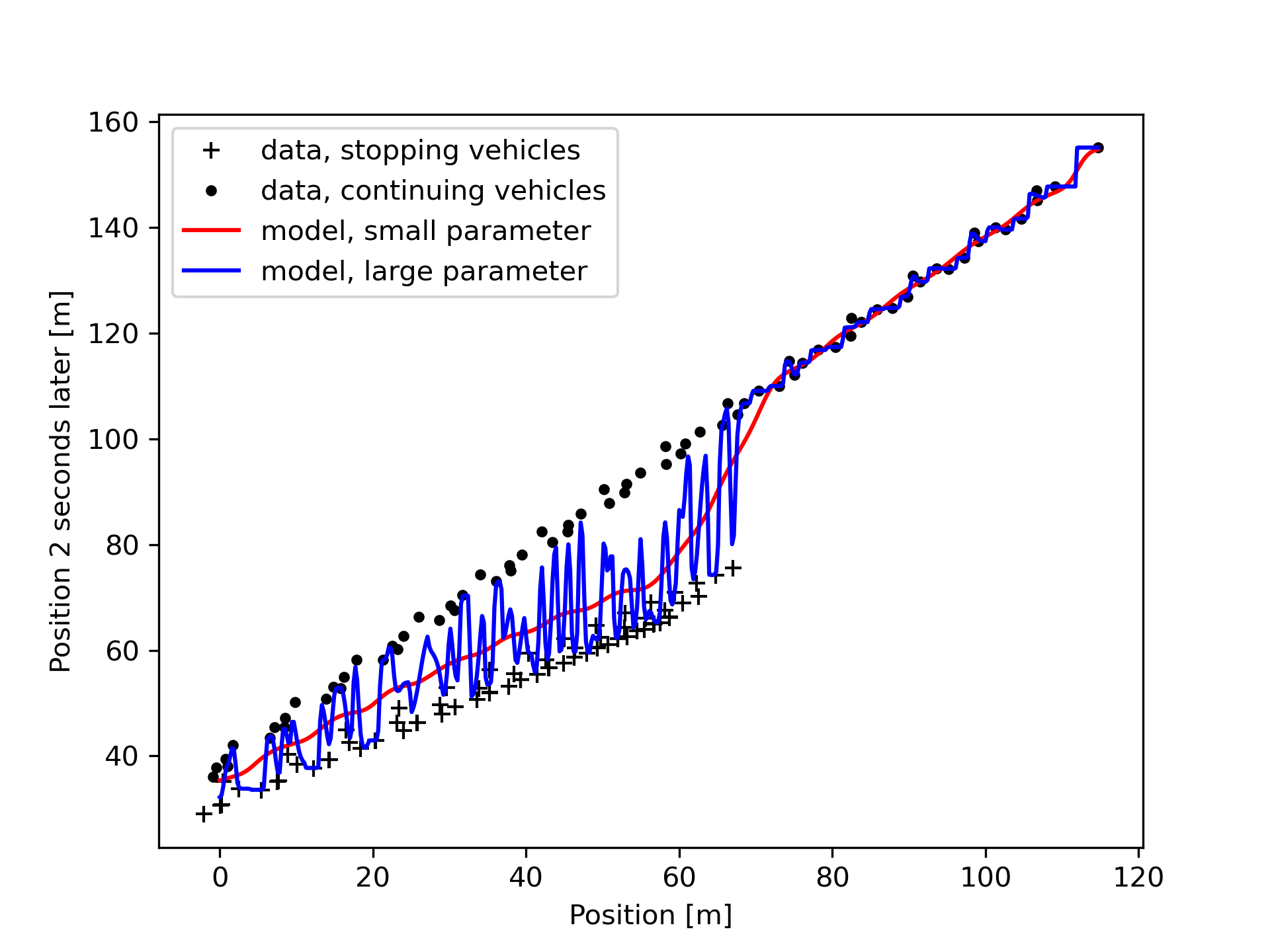}
				\label{fig:discussion:example_explaining_large_parameters}
			}
			\caption{}
		\end{figure*}

		Consider vehicles driving along a straight road. Assume that some vehicles deccelerate and stop while others keep their velocities. \autoref{fig:discussion:example_explaining_large_parameters_data} shows artificial data of such a scenario. This data is modeled like in \autoref{sec:method}, except that a road user state consists only of the current position of a vehicle, i.e.~$R = p$ and~$\bar{R}_{n,t} = \bar{p}_{n,t}$. \autoref{fig:discussion:example_explaining_large_parameters} shows how the model behaves when the parameters have small and large values. For some positions there is two possible future positions. Having large parameters makes the model behave like a nearest neighbor regression, which computes predictions based on few data points, making predictions jump between the two possible future positions. Having small parameters yields a prediction closer to the average of the two future positions. This is an example of an  ill-posed deterministic prediction problem, any deterministic prediction algorithm is expected to perform bad or demonstrate erratic predictions. In general, there is at least three options for proceeding:
		\begin{enumerate}
			\item restrict the domain of validity of the model, i.e. use the model for predictions only on data which demonstrate a well-posed deterministic prediction problem
			\item include, or change to, more suitable independent variables, i.e. model input, for determining the prediction
			\item formulate a probabilistic prediction model.
		\end{enumerate}
		The model in \autoref{sec:method} determines the future position from only current position, current speed and current orientation. Which intuitively is an ill-posed deterministic prediction problem.
	
	\subsection{An example of how to include road user interaction}\label{sec:discussion:interaction}
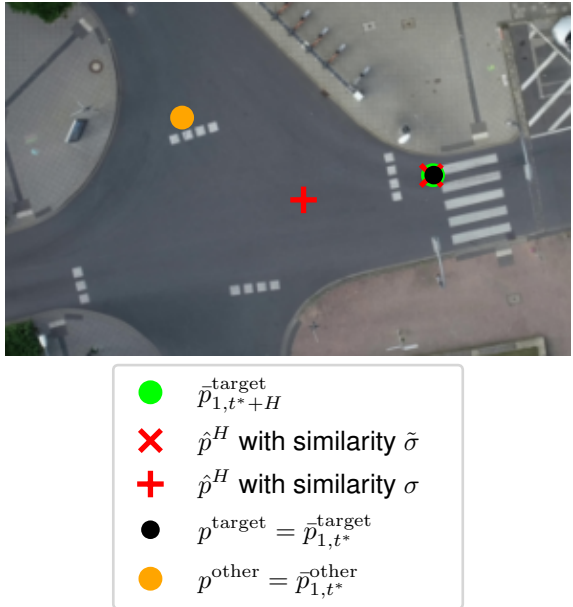
\begin{figure}
			\centering
			\input{figures/interaction_example.pgf}
			\caption{Prediction with and without using as input the other vehicle's state.}
			\label{fig:interaction_example}
		\end{figure}

\begin{figure}
			\centering
			\input{figures/interaction_example_distance.pgf}
			\caption{Prediction without using the other vehicle's state as input, i.e. with similarity $\sigma$.}
			\label{fig:interactino_example_distance}
		\end{figure}
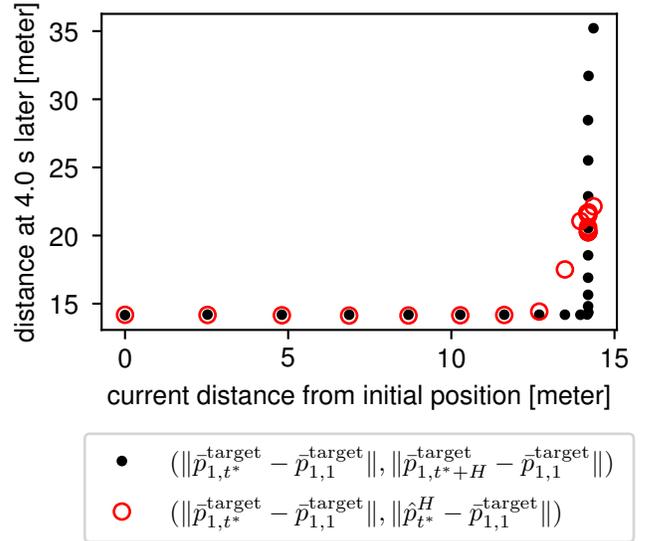

		\autoref{fig:interaction_example_locii_stops} shows positions of two vehicles as they approach an intersection, where one vehicle is by the law obliged to stop since the other vehicle approaches from its right hand side. Assume that the stopping vehicle is a target vehicle, whose future position should be predicted.
		The weighted average model in \autoref{sec:method} of the target vehicle neglects any effect a vehicle  close to the target vehicle has on it. 
		A weighted average model can be formulated to include such effects. In \autoref{sec:method} substitute ``road user" and ``target road user" with ``traffic situation" and ``currently observed traffic situation", respectively, and substitute~$R$ and~$\bar{R}_{n,t}$ with~$\mathcal{T}$ and~$\bar{\mathcal{T}}_{n,t}$. For example, if 
		\begin{enumerate}
			\item[(a)] the state of currently observed traffic situation is~\mbox{$\mathcal{T}= (p^\text{target},s,o,p^\text{other})$}, with~$p^\text{target}\in\mathbb{R}^2$, $s\in \mathbb{R}$ and $o\in \mathbb{R}^2$ as the current position, speed, and orientation of the target vehicle, as in \autoref{sec:method},  and~$p^\text{other}\in \mathbb{R}^2\cup\{None\}$ as the current position of the other vehicle, where $p^\text{other}$ assumes the value $None$ if the traffic situation lacks another vehicle than the target vehicle,
			\item[(b)] and $\bar{\mathcal{T}}_{n,t}$ with similarly defined,
			\item[(c)] and the similarity function is given by 
			\begin{equation}\label{eq:similarity_interaction}
				\begin{aligned}
					&\tilde{\sigma}(\mathcal{T},\bar{\mathcal{T}}_{n,t}) = \Big(\mathbb{I}_{\{None\}}(p^\text{other})\cdot \mathbb{I}_{\{None\}}(\bar{p}^\text{other}_{n,t})\\
					&\quad+(1-\mathbb{I}_{\{None\}}(p^\text{other})\cdot (1-\mathbb{I}_{\{None\}}(\bar{p}^\text{other}_{n,t}))\\
					&\quad\cdot \exp(\|p^\text{other}-\bar{p}^\text{other}_{n,t}\|^2)\Big)\\
					&\quad\cdot \sigma((p,s,o),(\bar{p}_{n,t},\bar{s}_{n,t},\bar{o}_{n,t}))
				\end{aligned}
			\end{equation}
			where $\sigma$ is as in \eqref{eq:similarity_metric}.
		\end{enumerate}
		then the resulting model includes effects that the ego-vehicle has on the target vehicle. Notice that the indicator functions amount to selecting from a dataset traffic situations where another vehicle is present when the target traffic situation has another vehicle, and conversely. The possible benefit of such a model is demonstrated on carefully selected real data. The selected data consists of two traffic situation state trajectories \mbox{$\{\bar{\mathcal{T}}_{1,t}\}_t=\{ (\bar{p}^\text{target}_{1,t},\bar{s}_{1,t},\bar{o}_{1,t},\bar{p}^\text{other}_{1,t})\}_t$} and \mbox{$\{\bar{\mathcal{T}}_{2,t}\}_t=\{ (\bar{p}^\text{target}_{2,t},\bar{s}_{2,t},\bar{o}_{2,t},\bar{p}^\text{other}_{2,t})\}_t$}. In the latter traffic situation another vehicle is lacking, thus $\forall t,\bar{p}^\text{other}_{2,t}=None$. The position trajectories of traffic situation trajectories $\{\bar{\mathcal{T}}_{1,t}\}_t$ and $\{\bar{\mathcal{T}}_{2,t}\}_t$ are shown in \autoref{fig:interaction_example_locii_stops} and \autoref{fig:interaction_example_locii_continues}, respectively. Trajectory $\{\bar{\mathcal{T}}_{1,t}\}_t$ is such that the target vehicle stops to give way to the other vehicle while trajectory $\{\bar{\mathcal{T}}_{1,t}\}_t$ is such that the target vehicle drives with approximately constant velocity. Both trajectories are designated as training data while trajectory $\{\bar{\mathcal{T}}_{2,t}\}_t$ is also designated as test data. 
		A time index $t^*$ is selected and it is assumed that the currently observed traffic situation state $\mathcal{T}$ is equal to the traffic situation state $\bar{\mathcal{T}}_{1,t^*}$. \autoref{fig:interaction_example} shows that evaluating model~\mbox{(a)-(c)} for a prediction horizon~$H= 4$ seconds results in sucessfully predicting the future position $\bar{\mathcal{T}}_{1,t^*+H}$, while using the model in \autoref{sec:method} fails. \autoref{fig:interactino_example_distance} illustrates the prediction of the model in \autoref{sec:method} as the $t^*$ varies. It is seen that the predictions are good up until the target vehicle starts waiting at the intersection, which is reasonable since the model has as input only the target vehicle's current position, speed and orientation, from which it is impossible to know \emph{when} the target vehicle starts moving again. On the other hand a similar plot for model (a)-(c) shows that this model successfully predicts the target vehicle's future positions. This is because model (a)-(c) includes as input the position of the other vehicle, which reveals when the target vehicle will start moving again. This indicates that in general, for a larger dataset and when training and test datasets are different, it is expected that including the position of another vehicle gives better performance than what is achieved by the model in \autoref{sec:method}.  Although model~\mbox{(a)-(b)} is sufficient for making this argument, the model may be too simple for more general data, since also the speed of the other vehicle may be relevant for determining when the target vehicle starts moving. Such inputs can be included by substituting $\exp(\|p^\text{other}-\bar{p}^\text{other}_{n,t}\|^2)$ in \eqref{eq:similarity_interaction} with $\exp(d\|p^\text{other}-\bar{p}^\text{other}_{n,t}\|^2+e|s^\text{other}-\bar{s}^\text{other}_{n,t}|^2)$, where $d$ and $e$ are parameters introduced for additional flexibility.

		\begin{figure*}
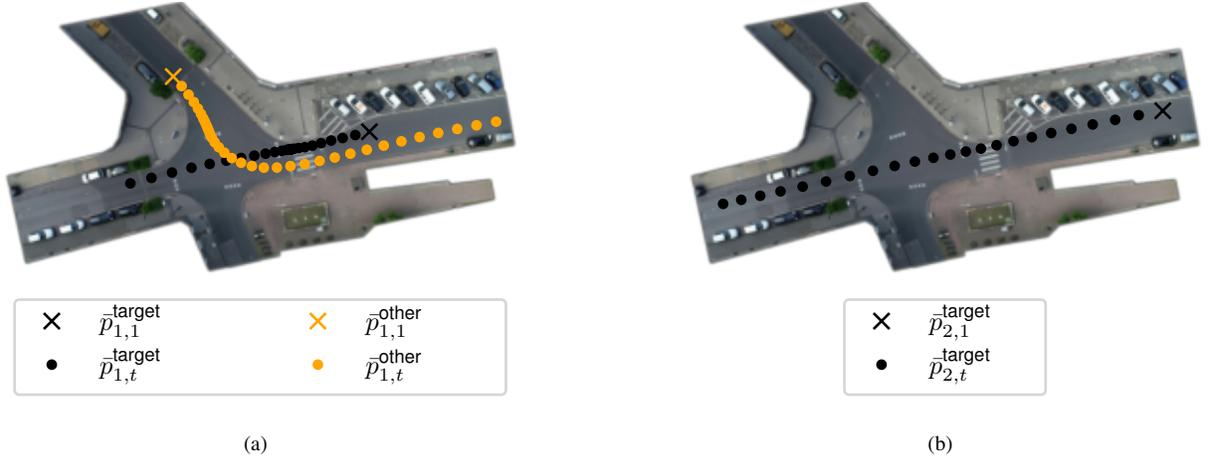

			\foreach \i/\case in {1/stops,2/continues}
				{
				\subfloat[ ]
						{
							\begin{minipage}{\columnwidth}
							\centering
							\begin{tikzpicture}
								\node (figure) {\input{figures/interaction_example_real_locii_\i.pgf}};
								\node[below=-0.45cm of figure] (legend) {\input{figures/interaction_example_real_locii_legend_\i.pgf}};
							\end{tikzpicture}%
							\end{minipage}%
							\label{fig:interaction_example_locii_\case}%
						}%
				}%
				\caption{Samples of from two artificially generated traffic situation state trajectories.}%
				\label{fig:interaction_example_locii}%
		\end{figure*}
		
	
	\subsection{Minimum prediction horizon for avoiding collisions}
		For evaluating the performance scores it is required to relate them to the braking distance of an intelligent vehicle performing predictions. An ideal prediction algorithm should perform well at an arbitrary horizon since it enables an intelligent vehicle to plan far ahead to both avoid collisions and ensure comfort for its passengers. On the other hand, if the application is emergency braking, then the minimum prediction horizon required for an intelligent vehicle to avoid a collision when travelling at some speed can be estimated from Newton's laws of motion and the standard model for friction. For an initial velocity of the vehicle, the minimum prediction horizon $H^\text{min}$ seconds must greater than or equal to the time it takes for the vehicle to come to rest while emergency braking. At an instant of time, the motion of  an emergency braking vehicle,  wheels locked, driving on a level surface and having mass $m$, acceleration $a$ and subject to earth's gravitational acceleration $g$, can be modeled such that for a coefficient of friction $\mu$ it is that the force on the vehicle is equal  to the force of the friction:
		\begin{equation}
			ma=-\mu mg.
		\end{equation}
		By imposing that the initial velocity of the vehicle is $v_0$ and that the velocity after $H^\text{min}$ seconds is zero and then solving the resulting equation for $H^\text{min}$ , gives that the minimum prediction horizon satisfies
		\begin{equation}
			H^\text{min}  = \frac{v_0}{\mu g}.
		\end{equation}
		The value of the coefficient of friction $\mu$ depends on factors such as tyre pressure, road type, and weather. According to \cite{wong2008} when a vehicle drives on dry or wet asphalt then $\mu=0.5$ and $\mu=0.8$, respectively. Hence, on dry apshalt the minimum prediction horizons for a vehicle moving at typical swedish city speeds of $30$ and $50$ km/h is $1.06$ and $1.77$ seconds, respectively, and on wet apshalt  $1.70$ and $2.83$ seconds. These estimates of the coefficient of friction are based on experiments done in the 1950s. With today's tires and modern anti-lock braking systems the minimum prediction horizon is expected to be less than the estimates given here. 
		
		

%% file: figures/interaction_example.pgf
\begingroup%
\makeatletter%
\begin{pgfpicture}%
\pgfpathrectangle{\pgfpointorigin}{\pgfqpoint{3.160000in}{3.375685in}}%
\pgfusepath{use as bounding box, clip}%
\begin{pgfscope}%
\pgfsetbuttcap%
\pgfsetmiterjoin%
\definecolor{currentfill}{rgb}{1.000000,1.000000,1.000000}%
\pgfsetfillcolor{currentfill}%
\pgfsetlinewidth{0.000000pt}%
\definecolor{currentstroke}{rgb}{1.000000,1.000000,1.000000}%
\pgfsetstrokecolor{currentstroke}%
\pgfsetdash{}{0pt}%
\pgfpathmoveto{\pgfqpoint{0.000000in}{0.000000in}}%
\pgfpathlineto{\pgfqpoint{3.160000in}{0.000000in}}%
\pgfpathlineto{\pgfqpoint{3.160000in}{3.375685in}}%
\pgfpathlineto{\pgfqpoint{0.000000in}{3.375685in}}%
\pgfpathclose%
\pgfusepath{fill}%
\end{pgfscope}%
\begin{pgfscope}%
\pgfsetbuttcap%
\pgfsetmiterjoin%
\definecolor{currentfill}{rgb}{1.000000,1.000000,1.000000}%
\pgfsetfillcolor{currentfill}%
\pgfsetlinewidth{0.000000pt}%
\definecolor{currentstroke}{rgb}{0.000000,0.000000,0.000000}%
\pgfsetstrokecolor{currentstroke}%
\pgfsetstrokeopacity{0.000000}%
\pgfsetdash{}{0pt}%
\pgfpathmoveto{\pgfqpoint{0.100000in}{1.425685in}}%
\pgfpathlineto{\pgfqpoint{3.060000in}{1.425685in}}%
\pgfpathlineto{\pgfqpoint{3.060000in}{3.275685in}}%
\pgfpathlineto{\pgfqpoint{0.100000in}{3.275685in}}%
\pgfpathclose%
\pgfusepath{fill}%
\end{pgfscope}%
\begin{pgfscope}%
\pgfpathrectangle{\pgfqpoint{0.100000in}{1.425685in}}{\pgfqpoint{2.960000in}{1.850000in}}%
\pgfusepath{clip}%
\pgfsys@transformshift{0.100000in}{1.425685in}%
\pgftext[left,bottom]{\includegraphics[interpolate=true,width=2.960000in,height=1.850000in]{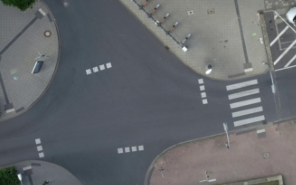}}%
\end{pgfscope}%
\begin{pgfscope}%
\pgfpathrectangle{\pgfqpoint{0.100000in}{1.425685in}}{\pgfqpoint{2.960000in}{1.850000in}}%
\pgfusepath{clip}%
\pgfsetbuttcap%
\pgfsetroundjoin%
\definecolor{currentfill}{rgb}{0.000000,1.000000,0.000000}%
\pgfsetfillcolor{currentfill}%
\pgfsetlinewidth{2.007500pt}%
\definecolor{currentstroke}{rgb}{0.000000,1.000000,0.000000}%
\pgfsetstrokecolor{currentstroke}%
\pgfsetdash{}{0pt}%
\pgfsys@defobject{currentmarker}{\pgfqpoint{-0.048611in}{-0.048611in}}{\pgfqpoint{0.048611in}{0.048611in}}{%
\pgfpathmoveto{\pgfqpoint{0.000000in}{-0.048611in}}%
\pgfpathcurveto{\pgfqpoint{0.012892in}{-0.048611in}}{\pgfqpoint{0.025257in}{-0.043489in}}{\pgfqpoint{0.034373in}{-0.034373in}}%
\pgfpathcurveto{\pgfqpoint{0.043489in}{-0.025257in}}{\pgfqpoint{0.048611in}{-0.012892in}}{\pgfqpoint{0.048611in}{0.000000in}}%
\pgfpathcurveto{\pgfqpoint{0.048611in}{0.012892in}}{\pgfqpoint{0.043489in}{0.025257in}}{\pgfqpoint{0.034373in}{0.034373in}}%
\pgfpathcurveto{\pgfqpoint{0.025257in}{0.043489in}}{\pgfqpoint{0.012892in}{0.048611in}}{\pgfqpoint{0.000000in}{0.048611in}}%
\pgfpathcurveto{\pgfqpoint{-0.012892in}{0.048611in}}{\pgfqpoint{-0.025257in}{0.043489in}}{\pgfqpoint{-0.034373in}{0.034373in}}%
\pgfpathcurveto{\pgfqpoint{-0.043489in}{0.025257in}}{\pgfqpoint{-0.048611in}{0.012892in}}{\pgfqpoint{-0.048611in}{0.000000in}}%
\pgfpathcurveto{\pgfqpoint{-0.048611in}{-0.012892in}}{\pgfqpoint{-0.043489in}{-0.025257in}}{\pgfqpoint{-0.034373in}{-0.034373in}}%
\pgfpathcurveto{\pgfqpoint{-0.025257in}{-0.043489in}}{\pgfqpoint{-0.012892in}{-0.048611in}}{\pgfqpoint{0.000000in}{-0.048611in}}%
\pgfpathclose%
\pgfusepath{stroke,fill}%
}%
\begin{pgfscope}%
\pgfsys@transformshift{2.338729in}{2.370933in}%
\pgfsys@useobject{currentmarker}{}%
\end{pgfscope}%
\end{pgfscope}%
\begin{pgfscope}%
\pgfpathrectangle{\pgfqpoint{0.100000in}{1.425685in}}{\pgfqpoint{2.960000in}{1.850000in}}%
\pgfusepath{clip}%
\pgfsetbuttcap%
\pgfsetroundjoin%
\definecolor{currentfill}{rgb}{1.000000,0.000000,0.000000}%
\pgfsetfillcolor{currentfill}%
\pgfsetlinewidth{2.007500pt}%
\definecolor{currentstroke}{rgb}{1.000000,0.000000,0.000000}%
\pgfsetstrokecolor{currentstroke}%
\pgfsetdash{}{0pt}%
\pgfsys@defobject{currentmarker}{\pgfqpoint{-0.055556in}{-0.055556in}}{\pgfqpoint{0.055556in}{0.055556in}}{%
\pgfpathmoveto{\pgfqpoint{-0.055556in}{-0.055556in}}%
\pgfpathlineto{\pgfqpoint{0.055556in}{0.055556in}}%
\pgfpathmoveto{\pgfqpoint{-0.055556in}{0.055556in}}%
\pgfpathlineto{\pgfqpoint{0.055556in}{-0.055556in}}%
\pgfusepath{stroke,fill}%
}%
\begin{pgfscope}%
\pgfsys@transformshift{2.330045in}{2.369773in}%
\pgfsys@useobject{currentmarker}{}%
\end{pgfscope}%
\end{pgfscope}%
\begin{pgfscope}%
\pgfpathrectangle{\pgfqpoint{0.100000in}{1.425685in}}{\pgfqpoint{2.960000in}{1.850000in}}%
\pgfusepath{clip}%
\pgfsetbuttcap%
\pgfsetroundjoin%
\definecolor{currentfill}{rgb}{1.000000,0.000000,0.000000}%
\pgfsetfillcolor{currentfill}%
\pgfsetlinewidth{2.007500pt}%
\definecolor{currentstroke}{rgb}{1.000000,0.000000,0.000000}%
\pgfsetstrokecolor{currentstroke}%
\pgfsetdash{}{0pt}%
\pgfsys@defobject{currentmarker}{\pgfqpoint{-0.069444in}{-0.069444in}}{\pgfqpoint{0.069444in}{0.069444in}}{%
\pgfpathmoveto{\pgfqpoint{-0.069444in}{0.000000in}}%
\pgfpathlineto{\pgfqpoint{0.069444in}{0.000000in}}%
\pgfpathmoveto{\pgfqpoint{0.000000in}{-0.069444in}}%
\pgfpathlineto{\pgfqpoint{0.000000in}{0.069444in}}%
\pgfusepath{stroke,fill}%
}%
\begin{pgfscope}%
\pgfsys@transformshift{1.661276in}{2.241852in}%
\pgfsys@useobject{currentmarker}{}%
\end{pgfscope}%
\end{pgfscope}%
\begin{pgfscope}%
\pgfpathrectangle{\pgfqpoint{0.100000in}{1.425685in}}{\pgfqpoint{2.960000in}{1.850000in}}%
\pgfusepath{clip}%
\pgfsetbuttcap%
\pgfsetroundjoin%
\definecolor{currentfill}{rgb}{0.000000,0.000000,0.000000}%
\pgfsetfillcolor{currentfill}%
\pgfsetlinewidth{1.003750pt}%
\definecolor{currentstroke}{rgb}{0.000000,0.000000,0.000000}%
\pgfsetstrokecolor{currentstroke}%
\pgfsetdash{}{0pt}%
\pgfsys@defobject{currentmarker}{\pgfqpoint{-0.041667in}{-0.041667in}}{\pgfqpoint{0.041667in}{0.041667in}}{%
\pgfpathmoveto{\pgfqpoint{0.000000in}{-0.041667in}}%
\pgfpathcurveto{\pgfqpoint{0.011050in}{-0.041667in}}{\pgfqpoint{0.021649in}{-0.037276in}}{\pgfqpoint{0.029463in}{-0.029463in}}%
\pgfpathcurveto{\pgfqpoint{0.037276in}{-0.021649in}}{\pgfqpoint{0.041667in}{-0.011050in}}{\pgfqpoint{0.041667in}{0.000000in}}%
\pgfpathcurveto{\pgfqpoint{0.041667in}{0.011050in}}{\pgfqpoint{0.037276in}{0.021649in}}{\pgfqpoint{0.029463in}{0.029463in}}%
\pgfpathcurveto{\pgfqpoint{0.021649in}{0.037276in}}{\pgfqpoint{0.011050in}{0.041667in}}{\pgfqpoint{0.000000in}{0.041667in}}%
\pgfpathcurveto{\pgfqpoint{-0.011050in}{0.041667in}}{\pgfqpoint{-0.021649in}{0.037276in}}{\pgfqpoint{-0.029463in}{0.029463in}}%
\pgfpathcurveto{\pgfqpoint{-0.037276in}{0.021649in}}{\pgfqpoint{-0.041667in}{0.011050in}}{\pgfqpoint{-0.041667in}{0.000000in}}%
\pgfpathcurveto{\pgfqpoint{-0.041667in}{-0.011050in}}{\pgfqpoint{-0.037276in}{-0.021649in}}{\pgfqpoint{-0.029463in}{-0.029463in}}%
\pgfpathcurveto{\pgfqpoint{-0.021649in}{-0.037276in}}{\pgfqpoint{-0.011050in}{-0.041667in}}{\pgfqpoint{0.000000in}{-0.041667in}}%
\pgfpathclose%
\pgfusepath{stroke,fill}%
}%
\begin{pgfscope}%
\pgfsys@transformshift{2.342430in}{2.371549in}%
\pgfsys@useobject{currentmarker}{}%
\end{pgfscope}%
\end{pgfscope}%
\begin{pgfscope}%
\pgfpathrectangle{\pgfqpoint{0.100000in}{1.425685in}}{\pgfqpoint{2.960000in}{1.850000in}}%
\pgfusepath{clip}%
\pgfsetbuttcap%
\pgfsetroundjoin%
\definecolor{currentfill}{rgb}{1.000000,0.647059,0.000000}%
\pgfsetfillcolor{currentfill}%
\pgfsetlinewidth{2.007500pt}%
\definecolor{currentstroke}{rgb}{1.000000,0.647059,0.000000}%
\pgfsetstrokecolor{currentstroke}%
\pgfsetdash{}{0pt}%
\pgfsys@defobject{currentmarker}{\pgfqpoint{-0.048611in}{-0.048611in}}{\pgfqpoint{0.048611in}{0.048611in}}{%
\pgfpathmoveto{\pgfqpoint{0.000000in}{-0.048611in}}%
\pgfpathcurveto{\pgfqpoint{0.012892in}{-0.048611in}}{\pgfqpoint{0.025257in}{-0.043489in}}{\pgfqpoint{0.034373in}{-0.034373in}}%
\pgfpathcurveto{\pgfqpoint{0.043489in}{-0.025257in}}{\pgfqpoint{0.048611in}{-0.012892in}}{\pgfqpoint{0.048611in}{0.000000in}}%
\pgfpathcurveto{\pgfqpoint{0.048611in}{0.012892in}}{\pgfqpoint{0.043489in}{0.025257in}}{\pgfqpoint{0.034373in}{0.034373in}}%
\pgfpathcurveto{\pgfqpoint{0.025257in}{0.043489in}}{\pgfqpoint{0.012892in}{0.048611in}}{\pgfqpoint{0.000000in}{0.048611in}}%
\pgfpathcurveto{\pgfqpoint{-0.012892in}{0.048611in}}{\pgfqpoint{-0.025257in}{0.043489in}}{\pgfqpoint{-0.034373in}{0.034373in}}%
\pgfpathcurveto{\pgfqpoint{-0.043489in}{0.025257in}}{\pgfqpoint{-0.048611in}{0.012892in}}{\pgfqpoint{-0.048611in}{0.000000in}}%
\pgfpathcurveto{\pgfqpoint{-0.048611in}{-0.012892in}}{\pgfqpoint{-0.043489in}{-0.025257in}}{\pgfqpoint{-0.034373in}{-0.034373in}}%
\pgfpathcurveto{\pgfqpoint{-0.025257in}{-0.043489in}}{\pgfqpoint{-0.012892in}{-0.048611in}}{\pgfqpoint{0.000000in}{-0.048611in}}%
\pgfpathclose%
\pgfusepath{stroke,fill}%
}%
\begin{pgfscope}%
\pgfsys@transformshift{1.025595in}{2.672869in}%
\pgfsys@useobject{currentmarker}{}%
\end{pgfscope}%
\end{pgfscope}%
\begin{pgfscope}%
\pgfsetbuttcap%
\pgfsetmiterjoin%
\definecolor{currentfill}{rgb}{1.000000,1.000000,1.000000}%
\pgfsetfillcolor{currentfill}%
\pgfsetfillopacity{0.800000}%
\pgfsetlinewidth{1.003750pt}%
\definecolor{currentstroke}{rgb}{0.800000,0.800000,0.800000}%
\pgfsetstrokecolor{currentstroke}%
\pgfsetstrokeopacity{0.800000}%
\pgfsetdash{}{0pt}%
\pgfpathmoveto{\pgfqpoint{0.693194in}{0.100000in}}%
\pgfpathlineto{\pgfqpoint{2.466806in}{0.100000in}}%
\pgfpathquadraticcurveto{\pgfqpoint{2.494584in}{0.100000in}}{\pgfqpoint{2.494584in}{0.127778in}}%
\pgfpathlineto{\pgfqpoint{2.494584in}{1.348593in}}%
\pgfpathquadraticcurveto{\pgfqpoint{2.494584in}{1.376371in}}{\pgfqpoint{2.466806in}{1.376371in}}%
\pgfpathlineto{\pgfqpoint{0.693194in}{1.376371in}}%
\pgfpathquadraticcurveto{\pgfqpoint{0.665416in}{1.376371in}}{\pgfqpoint{0.665416in}{1.348593in}}%
\pgfpathlineto{\pgfqpoint{0.665416in}{0.127778in}}%
\pgfpathquadraticcurveto{\pgfqpoint{0.665416in}{0.100000in}}{\pgfqpoint{0.693194in}{0.100000in}}%
\pgfpathclose%
\pgfusepath{stroke,fill}%
\end{pgfscope}%
\begin{pgfscope}%
\pgfsetbuttcap%
\pgfsetroundjoin%
\definecolor{currentfill}{rgb}{0.000000,1.000000,0.000000}%
\pgfsetfillcolor{currentfill}%
\pgfsetlinewidth{2.007500pt}%
\definecolor{currentstroke}{rgb}{0.000000,1.000000,0.000000}%
\pgfsetstrokecolor{currentstroke}%
\pgfsetdash{}{0pt}%
\pgfsys@defobject{currentmarker}{\pgfqpoint{-0.048611in}{-0.048611in}}{\pgfqpoint{0.048611in}{0.048611in}}{%
\pgfpathmoveto{\pgfqpoint{0.000000in}{-0.048611in}}%
\pgfpathcurveto{\pgfqpoint{0.012892in}{-0.048611in}}{\pgfqpoint{0.025257in}{-0.043489in}}{\pgfqpoint{0.034373in}{-0.034373in}}%
\pgfpathcurveto{\pgfqpoint{0.043489in}{-0.025257in}}{\pgfqpoint{0.048611in}{-0.012892in}}{\pgfqpoint{0.048611in}{0.000000in}}%
\pgfpathcurveto{\pgfqpoint{0.048611in}{0.012892in}}{\pgfqpoint{0.043489in}{0.025257in}}{\pgfqpoint{0.034373in}{0.034373in}}%
\pgfpathcurveto{\pgfqpoint{0.025257in}{0.043489in}}{\pgfqpoint{0.012892in}{0.048611in}}{\pgfqpoint{0.000000in}{0.048611in}}%
\pgfpathcurveto{\pgfqpoint{-0.012892in}{0.048611in}}{\pgfqpoint{-0.025257in}{0.043489in}}{\pgfqpoint{-0.034373in}{0.034373in}}%
\pgfpathcurveto{\pgfqpoint{-0.043489in}{0.025257in}}{\pgfqpoint{-0.048611in}{0.012892in}}{\pgfqpoint{-0.048611in}{0.000000in}}%
\pgfpathcurveto{\pgfqpoint{-0.048611in}{-0.012892in}}{\pgfqpoint{-0.043489in}{-0.025257in}}{\pgfqpoint{-0.034373in}{-0.034373in}}%
\pgfpathcurveto{\pgfqpoint{-0.025257in}{-0.043489in}}{\pgfqpoint{-0.012892in}{-0.048611in}}{\pgfqpoint{0.000000in}{-0.048611in}}%
\pgfpathclose%
\pgfusepath{stroke,fill}%
}%
\begin{pgfscope}%
\pgfsys@transformshift{0.859860in}{1.233859in}%
\pgfsys@useobject{currentmarker}{}%
\end{pgfscope}%
\end{pgfscope}%
\begin{pgfscope}%
\definecolor{textcolor}{rgb}{0.000000,0.000000,0.000000}%
\pgfsetstrokecolor{textcolor}%
\pgfsetfillcolor{textcolor}%
\pgftext[x=1.109860in,y=1.185248in,left,base]{\color{textcolor}\sffamily\fontsize{10.000000}{12.000000}\selectfont \(\displaystyle \bar{p}_{1,t^*+H}^\mathrm{target}\)}%
\end{pgfscope}%
\begin{pgfscope}%
\pgfsetbuttcap%
\pgfsetroundjoin%
\definecolor{currentfill}{rgb}{1.000000,0.000000,0.000000}%
\pgfsetfillcolor{currentfill}%
\pgfsetlinewidth{2.007500pt}%
\definecolor{currentstroke}{rgb}{1.000000,0.000000,0.000000}%
\pgfsetstrokecolor{currentstroke}%
\pgfsetdash{}{0pt}%
\pgfsys@defobject{currentmarker}{\pgfqpoint{-0.055556in}{-0.055556in}}{\pgfqpoint{0.055556in}{0.055556in}}{%
\pgfpathmoveto{\pgfqpoint{-0.055556in}{-0.055556in}}%
\pgfpathlineto{\pgfqpoint{0.055556in}{0.055556in}}%
\pgfpathmoveto{\pgfqpoint{-0.055556in}{0.055556in}}%
\pgfpathlineto{\pgfqpoint{0.055556in}{-0.055556in}}%
\pgfusepath{stroke,fill}%
}%
\begin{pgfscope}%
\pgfsys@transformshift{0.859860in}{0.980912in}%
\pgfsys@useobject{currentmarker}{}%
\end{pgfscope}%
\end{pgfscope}%
\begin{pgfscope}%
\definecolor{textcolor}{rgb}{0.000000,0.000000,0.000000}%
\pgfsetstrokecolor{textcolor}%
\pgfsetfillcolor{textcolor}%
\pgftext[x=1.109860in,y=0.932301in,left,base]{\color{textcolor}\sffamily\fontsize{10.000000}{12.000000}\selectfont \(\displaystyle \hat{p}^H\) with similarity \(\displaystyle \tilde{\sigma}\)}%
\end{pgfscope}%
\begin{pgfscope}%
\pgfsetbuttcap%
\pgfsetroundjoin%
\definecolor{currentfill}{rgb}{1.000000,0.000000,0.000000}%
\pgfsetfillcolor{currentfill}%
\pgfsetlinewidth{2.007500pt}%
\definecolor{currentstroke}{rgb}{1.000000,0.000000,0.000000}%
\pgfsetstrokecolor{currentstroke}%
\pgfsetdash{}{0pt}%
\pgfsys@defobject{currentmarker}{\pgfqpoint{-0.069444in}{-0.069444in}}{\pgfqpoint{0.069444in}{0.069444in}}{%
\pgfpathmoveto{\pgfqpoint{-0.069444in}{0.000000in}}%
\pgfpathlineto{\pgfqpoint{0.069444in}{0.000000in}}%
\pgfpathmoveto{\pgfqpoint{0.000000in}{-0.069444in}}%
\pgfpathlineto{\pgfqpoint{0.000000in}{0.069444in}}%
\pgfusepath{stroke,fill}%
}%
\begin{pgfscope}%
\pgfsys@transformshift{0.859860in}{0.753351in}%
\pgfsys@useobject{currentmarker}{}%
\end{pgfscope}%
\end{pgfscope}%
\begin{pgfscope}%
\definecolor{textcolor}{rgb}{0.000000,0.000000,0.000000}%
\pgfsetstrokecolor{textcolor}%
\pgfsetfillcolor{textcolor}%
\pgftext[x=1.109860in,y=0.704740in,left,base]{\color{textcolor}\sffamily\fontsize{10.000000}{12.000000}\selectfont \(\displaystyle \hat{p}^H\) with similarity \(\displaystyle \sigma\)}%
\end{pgfscope}%
\begin{pgfscope}%
\pgfsetbuttcap%
\pgfsetroundjoin%
\definecolor{currentfill}{rgb}{0.000000,0.000000,0.000000}%
\pgfsetfillcolor{currentfill}%
\pgfsetlinewidth{1.003750pt}%
\definecolor{currentstroke}{rgb}{0.000000,0.000000,0.000000}%
\pgfsetstrokecolor{currentstroke}%
\pgfsetdash{}{0pt}%
\pgfsys@defobject{currentmarker}{\pgfqpoint{-0.041667in}{-0.041667in}}{\pgfqpoint{0.041667in}{0.041667in}}{%
\pgfpathmoveto{\pgfqpoint{0.000000in}{-0.041667in}}%
\pgfpathcurveto{\pgfqpoint{0.011050in}{-0.041667in}}{\pgfqpoint{0.021649in}{-0.037276in}}{\pgfqpoint{0.029463in}{-0.029463in}}%
\pgfpathcurveto{\pgfqpoint{0.037276in}{-0.021649in}}{\pgfqpoint{0.041667in}{-0.011050in}}{\pgfqpoint{0.041667in}{0.000000in}}%
\pgfpathcurveto{\pgfqpoint{0.041667in}{0.011050in}}{\pgfqpoint{0.037276in}{0.021649in}}{\pgfqpoint{0.029463in}{0.029463in}}%
\pgfpathcurveto{\pgfqpoint{0.021649in}{0.037276in}}{\pgfqpoint{0.011050in}{0.041667in}}{\pgfqpoint{0.000000in}{0.041667in}}%
\pgfpathcurveto{\pgfqpoint{-0.011050in}{0.041667in}}{\pgfqpoint{-0.021649in}{0.037276in}}{\pgfqpoint{-0.029463in}{0.029463in}}%
\pgfpathcurveto{\pgfqpoint{-0.037276in}{0.021649in}}{\pgfqpoint{-0.041667in}{0.011050in}}{\pgfqpoint{-0.041667in}{0.000000in}}%
\pgfpathcurveto{\pgfqpoint{-0.041667in}{-0.011050in}}{\pgfqpoint{-0.037276in}{-0.021649in}}{\pgfqpoint{-0.029463in}{-0.029463in}}%
\pgfpathcurveto{\pgfqpoint{-0.021649in}{-0.037276in}}{\pgfqpoint{-0.011050in}{-0.041667in}}{\pgfqpoint{0.000000in}{-0.041667in}}%
\pgfpathclose%
\pgfusepath{stroke,fill}%
}%
\begin{pgfscope}%
\pgfsys@transformshift{0.859860in}{0.514003in}%
\pgfsys@useobject{currentmarker}{}%
\end{pgfscope}%
\end{pgfscope}%
\begin{pgfscope}%
\definecolor{textcolor}{rgb}{0.000000,0.000000,0.000000}%
\pgfsetstrokecolor{textcolor}%
\pgfsetfillcolor{textcolor}%
\pgftext[x=1.109860in,y=0.465392in,left,base]{\color{textcolor}\sffamily\fontsize{10.000000}{12.000000}\selectfont \(\displaystyle p^\mathrm{target}=\bar{p}_{1,t^*}^\mathrm{target}\)}%
\end{pgfscope}%
\begin{pgfscope}%
\pgfsetbuttcap%
\pgfsetroundjoin%
\definecolor{currentfill}{rgb}{1.000000,0.647059,0.000000}%
\pgfsetfillcolor{currentfill}%
\pgfsetlinewidth{2.007500pt}%
\definecolor{currentstroke}{rgb}{1.000000,0.647059,0.000000}%
\pgfsetstrokecolor{currentstroke}%
\pgfsetdash{}{0pt}%
\pgfsys@defobject{currentmarker}{\pgfqpoint{-0.048611in}{-0.048611in}}{\pgfqpoint{0.048611in}{0.048611in}}{%
\pgfpathmoveto{\pgfqpoint{0.000000in}{-0.048611in}}%
\pgfpathcurveto{\pgfqpoint{0.012892in}{-0.048611in}}{\pgfqpoint{0.025257in}{-0.043489in}}{\pgfqpoint{0.034373in}{-0.034373in}}%
\pgfpathcurveto{\pgfqpoint{0.043489in}{-0.025257in}}{\pgfqpoint{0.048611in}{-0.012892in}}{\pgfqpoint{0.048611in}{0.000000in}}%
\pgfpathcurveto{\pgfqpoint{0.048611in}{0.012892in}}{\pgfqpoint{0.043489in}{0.025257in}}{\pgfqpoint{0.034373in}{0.034373in}}%
\pgfpathcurveto{\pgfqpoint{0.025257in}{0.043489in}}{\pgfqpoint{0.012892in}{0.048611in}}{\pgfqpoint{0.000000in}{0.048611in}}%
\pgfpathcurveto{\pgfqpoint{-0.012892in}{0.048611in}}{\pgfqpoint{-0.025257in}{0.043489in}}{\pgfqpoint{-0.034373in}{0.034373in}}%
\pgfpathcurveto{\pgfqpoint{-0.043489in}{0.025257in}}{\pgfqpoint{-0.048611in}{0.012892in}}{\pgfqpoint{-0.048611in}{0.000000in}}%
\pgfpathcurveto{\pgfqpoint{-0.048611in}{-0.012892in}}{\pgfqpoint{-0.043489in}{-0.025257in}}{\pgfqpoint{-0.034373in}{-0.034373in}}%
\pgfpathcurveto{\pgfqpoint{-0.025257in}{-0.043489in}}{\pgfqpoint{-0.012892in}{-0.048611in}}{\pgfqpoint{0.000000in}{-0.048611in}}%
\pgfpathclose%
\pgfusepath{stroke,fill}%
}%
\begin{pgfscope}%
\pgfsys@transformshift{0.859860in}{0.257407in}%
\pgfsys@useobject{currentmarker}{}%
\end{pgfscope}%
\end{pgfscope}%
\begin{pgfscope}%
\definecolor{textcolor}{rgb}{0.000000,0.000000,0.000000}%
\pgfsetstrokecolor{textcolor}%
\pgfsetfillcolor{textcolor}%
\pgftext[x=1.109860in,y=0.208795in,left,base]{\color{textcolor}\sffamily\fontsize{10.000000}{12.000000}\selectfont \(\displaystyle p^\mathrm{other}=\bar{p}_{1,t^*}^\mathrm{other}\)}%
\end{pgfscope}%
\end{pgfpicture}%
\makeatother%
\endgroup%

%% file: figures/interaction_example_distance.pgf
\begingroup%
\makeatletter%
\begin{pgfpicture}%
\pgfpathrectangle{\pgfpointorigin}{\pgfqpoint{3.558272in}{3.183623in}}%
\pgfusepath{use as bounding box, clip}%
\begin{pgfscope}%
\pgfsetbuttcap%
\pgfsetmiterjoin%
\definecolor{currentfill}{rgb}{1.000000,1.000000,1.000000}%
\pgfsetfillcolor{currentfill}%
\pgfsetlinewidth{0.000000pt}%
\definecolor{currentstroke}{rgb}{1.000000,1.000000,1.000000}%
\pgfsetstrokecolor{currentstroke}%
\pgfsetdash{}{0pt}%
\pgfpathmoveto{\pgfqpoint{-0.000000in}{0.000000in}}%
\pgfpathlineto{\pgfqpoint{3.558272in}{0.000000in}}%
\pgfpathlineto{\pgfqpoint{3.558272in}{3.183623in}}%
\pgfpathlineto{\pgfqpoint{-0.000000in}{3.183623in}}%
\pgfpathclose%
\pgfusepath{fill}%
\end{pgfscope}%
\begin{pgfscope}%
\pgfsetbuttcap%
\pgfsetmiterjoin%
\definecolor{currentfill}{rgb}{1.000000,1.000000,1.000000}%
\pgfsetfillcolor{currentfill}%
\pgfsetlinewidth{0.000000pt}%
\definecolor{currentstroke}{rgb}{0.000000,0.000000,0.000000}%
\pgfsetstrokecolor{currentstroke}%
\pgfsetstrokeopacity{0.000000}%
\pgfsetdash{}{0pt}%
\pgfpathmoveto{\pgfqpoint{0.563921in}{1.211426in}}%
\pgfpathlineto{\pgfqpoint{3.260921in}{1.211426in}}%
\pgfpathlineto{\pgfqpoint{3.260921in}{2.866926in}}%
\pgfpathlineto{\pgfqpoint{0.563921in}{2.866926in}}%
\pgfpathclose%
\pgfusepath{fill}%
\end{pgfscope}%
\begin{pgfscope}%
\pgfsetbuttcap%
\pgfsetroundjoin%
\definecolor{currentfill}{rgb}{0.000000,0.000000,0.000000}%
\pgfsetfillcolor{currentfill}%
\pgfsetlinewidth{0.803000pt}%
\definecolor{currentstroke}{rgb}{0.000000,0.000000,0.000000}%
\pgfsetstrokecolor{currentstroke}%
\pgfsetdash{}{0pt}%
\pgfsys@defobject{currentmarker}{\pgfqpoint{0.000000in}{-0.048611in}}{\pgfqpoint{0.000000in}{0.000000in}}{%
\pgfpathmoveto{\pgfqpoint{0.000000in}{0.000000in}}%
\pgfpathlineto{\pgfqpoint{0.000000in}{-0.048611in}}%
\pgfusepath{stroke,fill}%
}%
\begin{pgfscope}%
\pgfsys@transformshift{0.686512in}{1.211426in}%
\pgfsys@useobject{currentmarker}{}%
\end{pgfscope}%
\end{pgfscope}%
\begin{pgfscope}%
\definecolor{textcolor}{rgb}{0.000000,0.000000,0.000000}%
\pgfsetstrokecolor{textcolor}%
\pgfsetfillcolor{textcolor}%
\pgftext[x=0.686512in,y=1.114203in,,top]{\color{textcolor}\sffamily\fontsize{10.000000}{12.000000}\selectfont 0}%
\end{pgfscope}%
\begin{pgfscope}%
\pgfsetbuttcap%
\pgfsetroundjoin%
\definecolor{currentfill}{rgb}{0.000000,0.000000,0.000000}%
\pgfsetfillcolor{currentfill}%
\pgfsetlinewidth{0.803000pt}%
\definecolor{currentstroke}{rgb}{0.000000,0.000000,0.000000}%
\pgfsetstrokecolor{currentstroke}%
\pgfsetdash{}{0pt}%
\pgfsys@defobject{currentmarker}{\pgfqpoint{0.000000in}{-0.048611in}}{\pgfqpoint{0.000000in}{0.000000in}}{%
\pgfpathmoveto{\pgfqpoint{0.000000in}{0.000000in}}%
\pgfpathlineto{\pgfqpoint{0.000000in}{-0.048611in}}%
\pgfusepath{stroke,fill}%
}%
\begin{pgfscope}%
\pgfsys@transformshift{1.540501in}{1.211426in}%
\pgfsys@useobject{currentmarker}{}%
\end{pgfscope}%
\end{pgfscope}%
\begin{pgfscope}%
\definecolor{textcolor}{rgb}{0.000000,0.000000,0.000000}%
\pgfsetstrokecolor{textcolor}%
\pgfsetfillcolor{textcolor}%
\pgftext[x=1.540501in,y=1.114203in,,top]{\color{textcolor}\sffamily\fontsize{10.000000}{12.000000}\selectfont 5}%
\end{pgfscope}%
\begin{pgfscope}%
\pgfsetbuttcap%
\pgfsetroundjoin%
\definecolor{currentfill}{rgb}{0.000000,0.000000,0.000000}%
\pgfsetfillcolor{currentfill}%
\pgfsetlinewidth{0.803000pt}%
\definecolor{currentstroke}{rgb}{0.000000,0.000000,0.000000}%
\pgfsetstrokecolor{currentstroke}%
\pgfsetdash{}{0pt}%
\pgfsys@defobject{currentmarker}{\pgfqpoint{0.000000in}{-0.048611in}}{\pgfqpoint{0.000000in}{0.000000in}}{%
\pgfpathmoveto{\pgfqpoint{0.000000in}{0.000000in}}%
\pgfpathlineto{\pgfqpoint{0.000000in}{-0.048611in}}%
\pgfusepath{stroke,fill}%
}%
\begin{pgfscope}%
\pgfsys@transformshift{2.394490in}{1.211426in}%
\pgfsys@useobject{currentmarker}{}%
\end{pgfscope}%
\end{pgfscope}%
\begin{pgfscope}%
\definecolor{textcolor}{rgb}{0.000000,0.000000,0.000000}%
\pgfsetstrokecolor{textcolor}%
\pgfsetfillcolor{textcolor}%
\pgftext[x=2.394490in,y=1.114203in,,top]{\color{textcolor}\sffamily\fontsize{10.000000}{12.000000}\selectfont 10}%
\end{pgfscope}%
\begin{pgfscope}%
\pgfsetbuttcap%
\pgfsetroundjoin%
\definecolor{currentfill}{rgb}{0.000000,0.000000,0.000000}%
\pgfsetfillcolor{currentfill}%
\pgfsetlinewidth{0.803000pt}%
\definecolor{currentstroke}{rgb}{0.000000,0.000000,0.000000}%
\pgfsetstrokecolor{currentstroke}%
\pgfsetdash{}{0pt}%
\pgfsys@defobject{currentmarker}{\pgfqpoint{0.000000in}{-0.048611in}}{\pgfqpoint{0.000000in}{0.000000in}}{%
\pgfpathmoveto{\pgfqpoint{0.000000in}{0.000000in}}%
\pgfpathlineto{\pgfqpoint{0.000000in}{-0.048611in}}%
\pgfusepath{stroke,fill}%
}%
\begin{pgfscope}%
\pgfsys@transformshift{3.248479in}{1.211426in}%
\pgfsys@useobject{currentmarker}{}%
\end{pgfscope}%
\end{pgfscope}%
\begin{pgfscope}%
\definecolor{textcolor}{rgb}{0.000000,0.000000,0.000000}%
\pgfsetstrokecolor{textcolor}%
\pgfsetfillcolor{textcolor}%
\pgftext[x=3.248479in,y=1.114203in,,top]{\color{textcolor}\sffamily\fontsize{10.000000}{12.000000}\selectfont 15}%
\end{pgfscope}%
\begin{pgfscope}%
\definecolor{textcolor}{rgb}{0.000000,0.000000,0.000000}%
\pgfsetstrokecolor{textcolor}%
\pgfsetfillcolor{textcolor}%
\pgftext[x=1.912421in,y=0.924235in,,top]{\color{textcolor}\sffamily\fontsize{10.000000}{12.000000}\selectfont current distance from initial position [meter]}%
\end{pgfscope}%
\begin{pgfscope}%
\pgfsetbuttcap%
\pgfsetroundjoin%
\definecolor{currentfill}{rgb}{0.000000,0.000000,0.000000}%
\pgfsetfillcolor{currentfill}%
\pgfsetlinewidth{0.803000pt}%
\definecolor{currentstroke}{rgb}{0.000000,0.000000,0.000000}%
\pgfsetstrokecolor{currentstroke}%
\pgfsetdash{}{0pt}%
\pgfsys@defobject{currentmarker}{\pgfqpoint{-0.048611in}{0.000000in}}{\pgfqpoint{-0.000000in}{0.000000in}}{%
\pgfpathmoveto{\pgfqpoint{-0.000000in}{0.000000in}}%
\pgfpathlineto{\pgfqpoint{-0.048611in}{0.000000in}}%
\pgfusepath{stroke,fill}%
}%
\begin{pgfscope}%
\pgfsys@transformshift{0.563921in}{1.348845in}%
\pgfsys@useobject{currentmarker}{}%
\end{pgfscope}%
\end{pgfscope}%
\begin{pgfscope}%
\definecolor{textcolor}{rgb}{0.000000,0.000000,0.000000}%
\pgfsetstrokecolor{textcolor}%
\pgfsetfillcolor{textcolor}%
\pgftext[x=0.289968in, y=1.296084in, left, base]{\color{textcolor}\sffamily\fontsize{10.000000}{12.000000}\selectfont 15}%
\end{pgfscope}%
\begin{pgfscope}%
\pgfsetbuttcap%
\pgfsetroundjoin%
\definecolor{currentfill}{rgb}{0.000000,0.000000,0.000000}%
\pgfsetfillcolor{currentfill}%
\pgfsetlinewidth{0.803000pt}%
\definecolor{currentstroke}{rgb}{0.000000,0.000000,0.000000}%
\pgfsetstrokecolor{currentstroke}%
\pgfsetdash{}{0pt}%
\pgfsys@defobject{currentmarker}{\pgfqpoint{-0.048611in}{0.000000in}}{\pgfqpoint{-0.000000in}{0.000000in}}{%
\pgfpathmoveto{\pgfqpoint{-0.000000in}{0.000000in}}%
\pgfpathlineto{\pgfqpoint{-0.048611in}{0.000000in}}%
\pgfusepath{stroke,fill}%
}%
\begin{pgfscope}%
\pgfsys@transformshift{0.563921in}{1.705769in}%
\pgfsys@useobject{currentmarker}{}%
\end{pgfscope}%
\end{pgfscope}%
\begin{pgfscope}%
\definecolor{textcolor}{rgb}{0.000000,0.000000,0.000000}%
\pgfsetstrokecolor{textcolor}%
\pgfsetfillcolor{textcolor}%
\pgftext[x=0.289968in, y=1.653008in, left, base]{\color{textcolor}\sffamily\fontsize{10.000000}{12.000000}\selectfont 20}%
\end{pgfscope}%
\begin{pgfscope}%
\pgfsetbuttcap%
\pgfsetroundjoin%
\definecolor{currentfill}{rgb}{0.000000,0.000000,0.000000}%
\pgfsetfillcolor{currentfill}%
\pgfsetlinewidth{0.803000pt}%
\definecolor{currentstroke}{rgb}{0.000000,0.000000,0.000000}%
\pgfsetstrokecolor{currentstroke}%
\pgfsetdash{}{0pt}%
\pgfsys@defobject{currentmarker}{\pgfqpoint{-0.048611in}{0.000000in}}{\pgfqpoint{-0.000000in}{0.000000in}}{%
\pgfpathmoveto{\pgfqpoint{-0.000000in}{0.000000in}}%
\pgfpathlineto{\pgfqpoint{-0.048611in}{0.000000in}}%
\pgfusepath{stroke,fill}%
}%
\begin{pgfscope}%
\pgfsys@transformshift{0.563921in}{2.062693in}%
\pgfsys@useobject{currentmarker}{}%
\end{pgfscope}%
\end{pgfscope}%
\begin{pgfscope}%
\definecolor{textcolor}{rgb}{0.000000,0.000000,0.000000}%
\pgfsetstrokecolor{textcolor}%
\pgfsetfillcolor{textcolor}%
\pgftext[x=0.289968in, y=2.009932in, left, base]{\color{textcolor}\sffamily\fontsize{10.000000}{12.000000}\selectfont 25}%
\end{pgfscope}%
\begin{pgfscope}%
\pgfsetbuttcap%
\pgfsetroundjoin%
\definecolor{currentfill}{rgb}{0.000000,0.000000,0.000000}%
\pgfsetfillcolor{currentfill}%
\pgfsetlinewidth{0.803000pt}%
\definecolor{currentstroke}{rgb}{0.000000,0.000000,0.000000}%
\pgfsetstrokecolor{currentstroke}%
\pgfsetdash{}{0pt}%
\pgfsys@defobject{currentmarker}{\pgfqpoint{-0.048611in}{0.000000in}}{\pgfqpoint{-0.000000in}{0.000000in}}{%
\pgfpathmoveto{\pgfqpoint{-0.000000in}{0.000000in}}%
\pgfpathlineto{\pgfqpoint{-0.048611in}{0.000000in}}%
\pgfusepath{stroke,fill}%
}%
\begin{pgfscope}%
\pgfsys@transformshift{0.563921in}{2.419617in}%
\pgfsys@useobject{currentmarker}{}%
\end{pgfscope}%
\end{pgfscope}%
\begin{pgfscope}%
\definecolor{textcolor}{rgb}{0.000000,0.000000,0.000000}%
\pgfsetstrokecolor{textcolor}%
\pgfsetfillcolor{textcolor}%
\pgftext[x=0.289968in, y=2.366855in, left, base]{\color{textcolor}\sffamily\fontsize{10.000000}{12.000000}\selectfont 30}%
\end{pgfscope}%
\begin{pgfscope}%
\pgfsetbuttcap%
\pgfsetroundjoin%
\definecolor{currentfill}{rgb}{0.000000,0.000000,0.000000}%
\pgfsetfillcolor{currentfill}%
\pgfsetlinewidth{0.803000pt}%
\definecolor{currentstroke}{rgb}{0.000000,0.000000,0.000000}%
\pgfsetstrokecolor{currentstroke}%
\pgfsetdash{}{0pt}%
\pgfsys@defobject{currentmarker}{\pgfqpoint{-0.048611in}{0.000000in}}{\pgfqpoint{-0.000000in}{0.000000in}}{%
\pgfpathmoveto{\pgfqpoint{-0.000000in}{0.000000in}}%
\pgfpathlineto{\pgfqpoint{-0.048611in}{0.000000in}}%
\pgfusepath{stroke,fill}%
}%
\begin{pgfscope}%
\pgfsys@transformshift{0.563921in}{2.776541in}%
\pgfsys@useobject{currentmarker}{}%
\end{pgfscope}%
\end{pgfscope}%
\begin{pgfscope}%
\definecolor{textcolor}{rgb}{0.000000,0.000000,0.000000}%
\pgfsetstrokecolor{textcolor}%
\pgfsetfillcolor{textcolor}%
\pgftext[x=0.289968in, y=2.723779in, left, base]{\color{textcolor}\sffamily\fontsize{10.000000}{12.000000}\selectfont 35}%
\end{pgfscope}%
\begin{pgfscope}%
\definecolor{textcolor}{rgb}{0.000000,0.000000,0.000000}%
\pgfsetstrokecolor{textcolor}%
\pgfsetfillcolor{textcolor}%
\pgftext[x=0.234413in,y=2.039176in,,bottom,rotate=90.000000]{\color{textcolor}\sffamily\fontsize{10.000000}{12.000000}\selectfont distance at 4.0 s later [meter]}%
\end{pgfscope}%
\begin{pgfscope}%
\pgfpathrectangle{\pgfqpoint{0.563921in}{1.211426in}}{\pgfqpoint{2.697000in}{1.655500in}}%
\pgfusepath{clip}%
\pgfsetbuttcap%
\pgfsetroundjoin%
\definecolor{currentfill}{rgb}{0.000000,0.000000,0.000000}%
\pgfsetfillcolor{currentfill}%
\pgfsetlinewidth{1.003750pt}%
\definecolor{currentstroke}{rgb}{0.000000,0.000000,0.000000}%
\pgfsetstrokecolor{currentstroke}%
\pgfsetdash{}{0pt}%
\pgfsys@defobject{currentmarker}{\pgfqpoint{-0.020833in}{-0.020833in}}{\pgfqpoint{0.020833in}{0.020833in}}{%
\pgfpathmoveto{\pgfqpoint{0.000000in}{-0.020833in}}%
\pgfpathcurveto{\pgfqpoint{0.005525in}{-0.020833in}}{\pgfqpoint{0.010825in}{-0.018638in}}{\pgfqpoint{0.014731in}{-0.014731in}}%
\pgfpathcurveto{\pgfqpoint{0.018638in}{-0.010825in}}{\pgfqpoint{0.020833in}{-0.005525in}}{\pgfqpoint{0.020833in}{0.000000in}}%
\pgfpathcurveto{\pgfqpoint{0.020833in}{0.005525in}}{\pgfqpoint{0.018638in}{0.010825in}}{\pgfqpoint{0.014731in}{0.014731in}}%
\pgfpathcurveto{\pgfqpoint{0.010825in}{0.018638in}}{\pgfqpoint{0.005525in}{0.020833in}}{\pgfqpoint{0.000000in}{0.020833in}}%
\pgfpathcurveto{\pgfqpoint{-0.005525in}{0.020833in}}{\pgfqpoint{-0.010825in}{0.018638in}}{\pgfqpoint{-0.014731in}{0.014731in}}%
\pgfpathcurveto{\pgfqpoint{-0.018638in}{0.010825in}}{\pgfqpoint{-0.020833in}{0.005525in}}{\pgfqpoint{-0.020833in}{0.000000in}}%
\pgfpathcurveto{\pgfqpoint{-0.020833in}{-0.005525in}}{\pgfqpoint{-0.018638in}{-0.010825in}}{\pgfqpoint{-0.014731in}{-0.014731in}}%
\pgfpathcurveto{\pgfqpoint{-0.010825in}{-0.018638in}}{\pgfqpoint{-0.005525in}{-0.020833in}}{\pgfqpoint{0.000000in}{-0.020833in}}%
\pgfpathclose%
\pgfusepath{stroke,fill}%
}%
\begin{pgfscope}%
\pgfsys@transformshift{0.686512in}{1.289092in}%
\pgfsys@useobject{currentmarker}{}%
\end{pgfscope}%
\begin{pgfscope}%
\pgfsys@transformshift{1.118185in}{1.291668in}%
\pgfsys@useobject{currentmarker}{}%
\end{pgfscope}%
\begin{pgfscope}%
\pgfsys@transformshift{1.507399in}{1.291077in}%
\pgfsys@useobject{currentmarker}{}%
\end{pgfscope}%
\begin{pgfscope}%
\pgfsys@transformshift{1.859483in}{1.291023in}%
\pgfsys@useobject{currentmarker}{}%
\end{pgfscope}%
\begin{pgfscope}%
\pgfsys@transformshift{2.170158in}{1.291058in}%
\pgfsys@useobject{currentmarker}{}%
\end{pgfscope}%
\begin{pgfscope}%
\pgfsys@transformshift{2.441221in}{1.291127in}%
\pgfsys@useobject{currentmarker}{}%
\end{pgfscope}%
\begin{pgfscope}%
\pgfsys@transformshift{2.671326in}{1.291158in}%
\pgfsys@useobject{currentmarker}{}%
\end{pgfscope}%
\begin{pgfscope}%
\pgfsys@transformshift{2.854447in}{1.291218in}%
\pgfsys@useobject{currentmarker}{}%
\end{pgfscope}%
\begin{pgfscope}%
\pgfsys@transformshift{2.988857in}{1.291191in}%
\pgfsys@useobject{currentmarker}{}%
\end{pgfscope}%
\begin{pgfscope}%
\pgfsys@transformshift{3.070022in}{1.290891in}%
\pgfsys@useobject{currentmarker}{}%
\end{pgfscope}%
\begin{pgfscope}%
\pgfsys@transformshift{3.105511in}{1.291983in}%
\pgfsys@useobject{currentmarker}{}%
\end{pgfscope}%
\begin{pgfscope}%
\pgfsys@transformshift{3.111674in}{1.302809in}%
\pgfsys@useobject{currentmarker}{}%
\end{pgfscope}%
\begin{pgfscope}%
\pgfsys@transformshift{3.110260in}{1.335479in}%
\pgfsys@useobject{currentmarker}{}%
\end{pgfscope}%
\begin{pgfscope}%
\pgfsys@transformshift{3.110132in}{1.395363in}%
\pgfsys@useobject{currentmarker}{}%
\end{pgfscope}%
\begin{pgfscope}%
\pgfsys@transformshift{3.110217in}{1.484942in}%
\pgfsys@useobject{currentmarker}{}%
\end{pgfscope}%
\begin{pgfscope}%
\pgfsys@transformshift{3.110381in}{1.602102in}%
\pgfsys@useobject{currentmarker}{}%
\end{pgfscope}%
\begin{pgfscope}%
\pgfsys@transformshift{3.110455in}{1.744566in}%
\pgfsys@useobject{currentmarker}{}%
\end{pgfscope}%
\begin{pgfscope}%
\pgfsys@transformshift{3.110599in}{1.911297in}%
\pgfsys@useobject{currentmarker}{}%
\end{pgfscope}%
\begin{pgfscope}%
\pgfsys@transformshift{3.110535in}{2.099240in}%
\pgfsys@useobject{currentmarker}{}%
\end{pgfscope}%
\begin{pgfscope}%
\pgfsys@transformshift{3.109816in}{2.310146in}%
\pgfsys@useobject{currentmarker}{}%
\end{pgfscope}%
\begin{pgfscope}%
\pgfsys@transformshift{3.112428in}{2.541943in}%
\pgfsys@useobject{currentmarker}{}%
\end{pgfscope}%
\begin{pgfscope}%
\pgfsys@transformshift{3.138330in}{2.791676in}%
\pgfsys@useobject{currentmarker}{}%
\end{pgfscope}%
\end{pgfscope}%
\begin{pgfscope}%
\pgfpathrectangle{\pgfqpoint{0.563921in}{1.211426in}}{\pgfqpoint{2.697000in}{1.655500in}}%
\pgfusepath{clip}%
\pgfsetbuttcap%
\pgfsetroundjoin%
\definecolor{currentfill}{rgb}{0.000000,0.000000,0.000000}%
\pgfsetfillcolor{currentfill}%
\pgfsetfillopacity{0.000000}%
\pgfsetlinewidth{1.003750pt}%
\definecolor{currentstroke}{rgb}{1.000000,0.000000,0.000000}%
\pgfsetstrokecolor{currentstroke}%
\pgfsetdash{}{0pt}%
\pgfsys@defobject{currentmarker}{\pgfqpoint{-0.041667in}{-0.041667in}}{\pgfqpoint{0.041667in}{0.041667in}}{%
\pgfpathmoveto{\pgfqpoint{0.000000in}{-0.041667in}}%
\pgfpathcurveto{\pgfqpoint{0.011050in}{-0.041667in}}{\pgfqpoint{0.021649in}{-0.037276in}}{\pgfqpoint{0.029463in}{-0.029463in}}%
\pgfpathcurveto{\pgfqpoint{0.037276in}{-0.021649in}}{\pgfqpoint{0.041667in}{-0.011050in}}{\pgfqpoint{0.041667in}{0.000000in}}%
\pgfpathcurveto{\pgfqpoint{0.041667in}{0.011050in}}{\pgfqpoint{0.037276in}{0.021649in}}{\pgfqpoint{0.029463in}{0.029463in}}%
\pgfpathcurveto{\pgfqpoint{0.021649in}{0.037276in}}{\pgfqpoint{0.011050in}{0.041667in}}{\pgfqpoint{0.000000in}{0.041667in}}%
\pgfpathcurveto{\pgfqpoint{-0.011050in}{0.041667in}}{\pgfqpoint{-0.021649in}{0.037276in}}{\pgfqpoint{-0.029463in}{0.029463in}}%
\pgfpathcurveto{\pgfqpoint{-0.037276in}{0.021649in}}{\pgfqpoint{-0.041667in}{0.011050in}}{\pgfqpoint{-0.041667in}{0.000000in}}%
\pgfpathcurveto{\pgfqpoint{-0.041667in}{-0.011050in}}{\pgfqpoint{-0.037276in}{-0.021649in}}{\pgfqpoint{-0.029463in}{-0.029463in}}%
\pgfpathcurveto{\pgfqpoint{-0.021649in}{-0.037276in}}{\pgfqpoint{-0.011050in}{-0.041667in}}{\pgfqpoint{0.000000in}{-0.041667in}}%
\pgfpathclose%
\pgfusepath{stroke,fill}%
}%
\begin{pgfscope}%
\pgfsys@transformshift{0.686512in}{1.290551in}%
\pgfsys@useobject{currentmarker}{}%
\end{pgfscope}%
\begin{pgfscope}%
\pgfsys@transformshift{1.118185in}{1.289936in}%
\pgfsys@useobject{currentmarker}{}%
\end{pgfscope}%
\begin{pgfscope}%
\pgfsys@transformshift{1.507399in}{1.288094in}%
\pgfsys@useobject{currentmarker}{}%
\end{pgfscope}%
\begin{pgfscope}%
\pgfsys@transformshift{1.859483in}{1.286676in}%
\pgfsys@useobject{currentmarker}{}%
\end{pgfscope}%
\begin{pgfscope}%
\pgfsys@transformshift{2.170158in}{1.287019in}%
\pgfsys@useobject{currentmarker}{}%
\end{pgfscope}%
\begin{pgfscope}%
\pgfsys@transformshift{2.441221in}{1.288108in}%
\pgfsys@useobject{currentmarker}{}%
\end{pgfscope}%
\begin{pgfscope}%
\pgfsys@transformshift{2.671326in}{1.289660in}%
\pgfsys@useobject{currentmarker}{}%
\end{pgfscope}%
\begin{pgfscope}%
\pgfsys@transformshift{2.854447in}{1.307487in}%
\pgfsys@useobject{currentmarker}{}%
\end{pgfscope}%
\begin{pgfscope}%
\pgfsys@transformshift{2.988857in}{1.528177in}%
\pgfsys@useobject{currentmarker}{}%
\end{pgfscope}%
\begin{pgfscope}%
\pgfsys@transformshift{3.070022in}{1.781759in}%
\pgfsys@useobject{currentmarker}{}%
\end{pgfscope}%
\begin{pgfscope}%
\pgfsys@transformshift{3.105511in}{1.824022in}%
\pgfsys@useobject{currentmarker}{}%
\end{pgfscope}%
\begin{pgfscope}%
\pgfsys@transformshift{3.111674in}{1.812308in}%
\pgfsys@useobject{currentmarker}{}%
\end{pgfscope}%
\begin{pgfscope}%
\pgfsys@transformshift{3.110260in}{1.723046in}%
\pgfsys@useobject{currentmarker}{}%
\end{pgfscope}%
\begin{pgfscope}%
\pgfsys@transformshift{3.110132in}{1.724268in}%
\pgfsys@useobject{currentmarker}{}%
\end{pgfscope}%
\begin{pgfscope}%
\pgfsys@transformshift{3.110217in}{1.724296in}%
\pgfsys@useobject{currentmarker}{}%
\end{pgfscope}%
\begin{pgfscope}%
\pgfsys@transformshift{3.110381in}{1.724365in}%
\pgfsys@useobject{currentmarker}{}%
\end{pgfscope}%
\begin{pgfscope}%
\pgfsys@transformshift{3.110455in}{1.724395in}%
\pgfsys@useobject{currentmarker}{}%
\end{pgfscope}%
\begin{pgfscope}%
\pgfsys@transformshift{3.110599in}{1.724459in}%
\pgfsys@useobject{currentmarker}{}%
\end{pgfscope}%
\begin{pgfscope}%
\pgfsys@transformshift{3.110535in}{1.739931in}%
\pgfsys@useobject{currentmarker}{}%
\end{pgfscope}%
\begin{pgfscope}%
\pgfsys@transformshift{3.109816in}{1.750617in}%
\pgfsys@useobject{currentmarker}{}%
\end{pgfscope}%
\begin{pgfscope}%
\pgfsys@transformshift{3.112428in}{1.826926in}%
\pgfsys@useobject{currentmarker}{}%
\end{pgfscope}%
\begin{pgfscope}%
\pgfsys@transformshift{3.138330in}{1.858719in}%
\pgfsys@useobject{currentmarker}{}%
\end{pgfscope}%
\end{pgfscope}%
\begin{pgfscope}%
\pgfsetrectcap%
\pgfsetmiterjoin%
\pgfsetlinewidth{0.803000pt}%
\definecolor{currentstroke}{rgb}{0.000000,0.000000,0.000000}%
\pgfsetstrokecolor{currentstroke}%
\pgfsetdash{}{0pt}%
\pgfpathmoveto{\pgfqpoint{0.563921in}{1.211426in}}%
\pgfpathlineto{\pgfqpoint{0.563921in}{2.866926in}}%
\pgfusepath{stroke}%
\end{pgfscope}%
\begin{pgfscope}%
\pgfsetrectcap%
\pgfsetmiterjoin%
\pgfsetlinewidth{0.803000pt}%
\definecolor{currentstroke}{rgb}{0.000000,0.000000,0.000000}%
\pgfsetstrokecolor{currentstroke}%
\pgfsetdash{}{0pt}%
\pgfpathmoveto{\pgfqpoint{3.260921in}{1.211426in}}%
\pgfpathlineto{\pgfqpoint{3.260921in}{2.866926in}}%
\pgfusepath{stroke}%
\end{pgfscope}%
\begin{pgfscope}%
\pgfsetrectcap%
\pgfsetmiterjoin%
\pgfsetlinewidth{0.803000pt}%
\definecolor{currentstroke}{rgb}{0.000000,0.000000,0.000000}%
\pgfsetstrokecolor{currentstroke}%
\pgfsetdash{}{0pt}%
\pgfpathmoveto{\pgfqpoint{0.563921in}{1.211426in}}%
\pgfpathlineto{\pgfqpoint{3.260921in}{1.211426in}}%
\pgfusepath{stroke}%
\end{pgfscope}%
\begin{pgfscope}%
\pgfsetrectcap%
\pgfsetmiterjoin%
\pgfsetlinewidth{0.803000pt}%
\definecolor{currentstroke}{rgb}{0.000000,0.000000,0.000000}%
\pgfsetstrokecolor{currentstroke}%
\pgfsetdash{}{0pt}%
\pgfpathmoveto{\pgfqpoint{0.563921in}{2.866926in}}%
\pgfpathlineto{\pgfqpoint{3.260921in}{2.866926in}}%
\pgfusepath{stroke}%
\end{pgfscope}%
\begin{pgfscope}%
\pgfsetbuttcap%
\pgfsetmiterjoin%
\definecolor{currentfill}{rgb}{1.000000,1.000000,1.000000}%
\pgfsetfillcolor{currentfill}%
\pgfsetfillopacity{0.800000}%
\pgfsetlinewidth{1.003750pt}%
\definecolor{currentstroke}{rgb}{0.800000,0.800000,0.800000}%
\pgfsetstrokecolor{currentstroke}%
\pgfsetstrokeopacity{0.800000}%
\pgfsetdash{}{0pt}%
\pgfpathmoveto{\pgfqpoint{0.503498in}{0.100000in}}%
\pgfpathlineto{\pgfqpoint{3.321344in}{0.100000in}}%
\pgfpathquadraticcurveto{\pgfqpoint{3.349122in}{0.100000in}}{\pgfqpoint{3.349122in}{0.127778in}}%
\pgfpathlineto{\pgfqpoint{3.349122in}{0.639573in}}%
\pgfpathquadraticcurveto{\pgfqpoint{3.349122in}{0.667351in}}{\pgfqpoint{3.321344in}{0.667351in}}%
\pgfpathlineto{\pgfqpoint{0.503498in}{0.667351in}}%
\pgfpathquadraticcurveto{\pgfqpoint{0.475721in}{0.667351in}}{\pgfqpoint{0.475721in}{0.639573in}}%
\pgfpathlineto{\pgfqpoint{0.475721in}{0.127778in}}%
\pgfpathquadraticcurveto{\pgfqpoint{0.475721in}{0.100000in}}{\pgfqpoint{0.503498in}{0.100000in}}%
\pgfpathclose%
\pgfusepath{stroke,fill}%
\end{pgfscope}%
\begin{pgfscope}%
\pgfsetbuttcap%
\pgfsetroundjoin%
\definecolor{currentfill}{rgb}{0.000000,0.000000,0.000000}%
\pgfsetfillcolor{currentfill}%
\pgfsetlinewidth{1.003750pt}%
\definecolor{currentstroke}{rgb}{0.000000,0.000000,0.000000}%
\pgfsetstrokecolor{currentstroke}%
\pgfsetdash{}{0pt}%
\pgfsys@defobject{currentmarker}{\pgfqpoint{-0.020833in}{-0.020833in}}{\pgfqpoint{0.020833in}{0.020833in}}{%
\pgfpathmoveto{\pgfqpoint{0.000000in}{-0.020833in}}%
\pgfpathcurveto{\pgfqpoint{0.005525in}{-0.020833in}}{\pgfqpoint{0.010825in}{-0.018638in}}{\pgfqpoint{0.014731in}{-0.014731in}}%
\pgfpathcurveto{\pgfqpoint{0.018638in}{-0.010825in}}{\pgfqpoint{0.020833in}{-0.005525in}}{\pgfqpoint{0.020833in}{0.000000in}}%
\pgfpathcurveto{\pgfqpoint{0.020833in}{0.005525in}}{\pgfqpoint{0.018638in}{0.010825in}}{\pgfqpoint{0.014731in}{0.014731in}}%
\pgfpathcurveto{\pgfqpoint{0.010825in}{0.018638in}}{\pgfqpoint{0.005525in}{0.020833in}}{\pgfqpoint{0.000000in}{0.020833in}}%
\pgfpathcurveto{\pgfqpoint{-0.005525in}{0.020833in}}{\pgfqpoint{-0.010825in}{0.018638in}}{\pgfqpoint{-0.014731in}{0.014731in}}%
\pgfpathcurveto{\pgfqpoint{-0.018638in}{0.010825in}}{\pgfqpoint{-0.020833in}{0.005525in}}{\pgfqpoint{-0.020833in}{0.000000in}}%
\pgfpathcurveto{\pgfqpoint{-0.020833in}{-0.005525in}}{\pgfqpoint{-0.018638in}{-0.010825in}}{\pgfqpoint{-0.014731in}{-0.014731in}}%
\pgfpathcurveto{\pgfqpoint{-0.010825in}{-0.018638in}}{\pgfqpoint{-0.005525in}{-0.020833in}}{\pgfqpoint{0.000000in}{-0.020833in}}%
\pgfpathclose%
\pgfusepath{stroke,fill}%
}%
\begin{pgfscope}%
\pgfsys@transformshift{0.670165in}{0.524840in}%
\pgfsys@useobject{currentmarker}{}%
\end{pgfscope}%
\end{pgfscope}%
\begin{pgfscope}%
\definecolor{textcolor}{rgb}{0.000000,0.000000,0.000000}%
\pgfsetstrokecolor{textcolor}%
\pgfsetfillcolor{textcolor}%
\pgftext[x=0.920165in,y=0.476229in,left,base]{\color{textcolor}\sffamily\fontsize{10.000000}{12.000000}\selectfont \(\displaystyle (\|\bar{p}_{1,t^*}^\mathrm{target}-\bar{p}_{1,1}^\mathrm{target}\|,\|\bar{p}_{1,t^*+H}^\mathrm{target}-\bar{p}_{1,1}^\mathrm{target}\|)\)}%
\end{pgfscope}%
\begin{pgfscope}%
\pgfsetbuttcap%
\pgfsetroundjoin%
\definecolor{currentfill}{rgb}{0.000000,0.000000,0.000000}%
\pgfsetfillcolor{currentfill}%
\pgfsetfillopacity{0.000000}%
\pgfsetlinewidth{1.003750pt}%
\definecolor{currentstroke}{rgb}{1.000000,0.000000,0.000000}%
\pgfsetstrokecolor{currentstroke}%
\pgfsetdash{}{0pt}%
\pgfsys@defobject{currentmarker}{\pgfqpoint{-0.041667in}{-0.041667in}}{\pgfqpoint{0.041667in}{0.041667in}}{%
\pgfpathmoveto{\pgfqpoint{0.000000in}{-0.041667in}}%
\pgfpathcurveto{\pgfqpoint{0.011050in}{-0.041667in}}{\pgfqpoint{0.021649in}{-0.037276in}}{\pgfqpoint{0.029463in}{-0.029463in}}%
\pgfpathcurveto{\pgfqpoint{0.037276in}{-0.021649in}}{\pgfqpoint{0.041667in}{-0.011050in}}{\pgfqpoint{0.041667in}{0.000000in}}%
\pgfpathcurveto{\pgfqpoint{0.041667in}{0.011050in}}{\pgfqpoint{0.037276in}{0.021649in}}{\pgfqpoint{0.029463in}{0.029463in}}%
\pgfpathcurveto{\pgfqpoint{0.021649in}{0.037276in}}{\pgfqpoint{0.011050in}{0.041667in}}{\pgfqpoint{0.000000in}{0.041667in}}%
\pgfpathcurveto{\pgfqpoint{-0.011050in}{0.041667in}}{\pgfqpoint{-0.021649in}{0.037276in}}{\pgfqpoint{-0.029463in}{0.029463in}}%
\pgfpathcurveto{\pgfqpoint{-0.037276in}{0.021649in}}{\pgfqpoint{-0.041667in}{0.011050in}}{\pgfqpoint{-0.041667in}{0.000000in}}%
\pgfpathcurveto{\pgfqpoint{-0.041667in}{-0.011050in}}{\pgfqpoint{-0.037276in}{-0.021649in}}{\pgfqpoint{-0.029463in}{-0.029463in}}%
\pgfpathcurveto{\pgfqpoint{-0.021649in}{-0.037276in}}{\pgfqpoint{-0.011050in}{-0.041667in}}{\pgfqpoint{0.000000in}{-0.041667in}}%
\pgfpathclose%
\pgfusepath{stroke,fill}%
}%
\begin{pgfscope}%
\pgfsys@transformshift{0.670165in}{0.260107in}%
\pgfsys@useobject{currentmarker}{}%
\end{pgfscope}%
\end{pgfscope}%
\begin{pgfscope}%
\definecolor{textcolor}{rgb}{0.000000,0.000000,0.000000}%
\pgfsetstrokecolor{textcolor}%
\pgfsetfillcolor{textcolor}%
\pgftext[x=0.920165in,y=0.211496in,left,base]{\color{textcolor}\sffamily\fontsize{10.000000}{12.000000}\selectfont \(\displaystyle (\|\bar{p}_{1,t^*}^\mathrm{target}-\bar{p}_{1,1}^\mathrm{target}\|,\|\hat{p}^H_{t^*}-\bar{p}_{1,1}^\mathrm{target}\|)\)}%
\end{pgfscope}%
\end{pgfpicture}%
\makeatother%
\endgroup%

%% file: conclusion.tex
\section{Conclusion}
	A comparison between our formulation of road user position prediction as a weighted average and previously reported state-of-the-art results on real world data shows that further improvement of the formulation is required. This may be explained by the state-of-the-art algorithms including as inputs the positions of every road user surrounding the target road user and including as input also past samples of positions, while our formulation only has as input the target road user's current position, speed, and orientation. Therefore an example of how to re-formulate our model to include such inputs was presented. The example shows that such input can be included without significantly scaling the complexity of our method. Still, the re-formulation requires selecting from all road users surrounding the target road user the road users that influence the target road user the most, where one possible model of influence is via oppinion dynamics \cite{heiker2022}. On the other hand, the state-of-the-arts are based on complex neural networks that have many more, some even with millions more, and less interpretable parameters than the weighted average model. The many parameters of neural networks provide great flexibility for successful application on many types of problems, but their flexibility is also a weakness as demonstrated in \cite{su2019one}, where the outputs of neural network image classification algorithms are significantly altered by changing a single pixel in the input images. More examples of such unsafe behavior is found in~\cite{miller2020adversarial,kong2021survey,huang2020survey}.
	
	Experimental results on real world data indicate that our formulation of road user position prediction as a weighted average produces $H$-second prediction errors comparable to that of the baseline constant velocity model at prediction horizons~$H$ smaller than approximately~$1.6$ seconds for vehicles, bicycles and pedestrians, which indicates that the model needs further adaptations. Hence, if a prediction model is sought for the application of emergency braking in urban environments, speeds less than~$30$ km/h, then the constant velocity model is preferable due to its simplicity. 
	On the other hand, if a prediction model is meant to support motion planning for autonomous driving applications, where the planned motion is required to be safe and smooth, 
	then longer prediction horizons are required and the weighted average model is preferable over the constant velocity model. 
	
	In most cases the weighted average model and a baseline neural network model performs comparably well. For the case of pedestrian position prediction the weighted average model tends to outperform the neural network baseline, while for vehicles the relationship is reversed.
	Still, our model has only~$3$ parameters, with clear interpretations, against the thousands of parameters of the neural network model, which are also more difficult to analyze and interpret.

	 The variation of the~$H$-second prediction errors of the weighted average model should be studied as the amount of data increases. As increasing the amount of data used by the model increases the online computational load, techniques for pre-processing newly available data should be explored in order to balance performance improvements and computational load.
	 
	 If a particular road or side walk in a dataset has few trajectories of road users traversing it, then techniques for automatically switching from the weighted average model to a more suitable model may be required, since the weighted average model performs less well when there are few examples.
	 
	 In the weighted average model, weights are interpreted as similarity between two road users. It is imposed that similarity decays monotonically as the road users become more different. While we have studied exponential rate of decay, other rates could improve performance. We defined similarity as a comparison between two road users' current position, speed and orientations, which are factors that do not relate to a unique future position. Including factors based on surrounding road users\cite{martin2020}, the geometry of a location\cite{karasev2016} and traffic laws\cite{koschi2020} could further improve the model. Currently, our method lacks an automatic way of detecting which values of used factors fail to relate to a unique future position, incorporating this could require a probabilistic formulation.

%% file: appendix.tex
\appendices
\section{Proof of weighted average} \label{sec:appendix:proof}
	Express the objective in \eqref{eq:methodology:optimality_formulation} as
	\begin{equation}
		\sum_{(\bar{R},\bar{d})\in X} \sigma(R,\bar{R})(d-\bar{d})^\top (d - \bar{d}).
	\end{equation}
	By \cite[Eq. 83]{petersen2012matrix} the derivative with respect to $d$ is 
	\begin{equation}
		\sum_{(\bar{R},\bar{d})\in X} 2\sigma(R,\bar{R})(d-\bar{d})
	\end{equation}
	Equating to zero and solving for $d$ gives \eqref{eq:methodology:nadaraya-watson}.